\documentclass{article} 
\usepackage{iclr2026_conference,times}

\usepackage{amsmath,amsfonts,bm}

\def\eqref#1{equation~\ref{#1}}

\def\1{\bm{1}}

\DeclareMathAlphabet{\mathsfit}{\encodingdefault}{\sfdefault}{m}{sl}
\SetMathAlphabet{\mathsfit}{bold}{\encodingdefault}{\sfdefault}{bx}{n}

\usepackage{hyperref}
\usepackage{url}
\usepackage[utf8]{inputenc} 
\usepackage[T1]{fontenc}    
\usepackage{booktabs}       
\usepackage{amsfonts}       
\usepackage{nicefrac}       
\usepackage{microtype}      
\usepackage{xcolor}         
\usepackage{graphicx}
\usepackage{multirow}
\usepackage[table]{xcolor} 
\usepackage{wrapfig}
\usepackage{pifont}
\usepackage{amsmath}
\usepackage{makecell}  

\title{LaVCa: LLM-assisted Visual Cortex Captioning}

\author{Takuya Matsuyama$^{1,2}$ \\
\texttt{u722117c@ecs.osaka-u.ac.jp}
\And
Shinji Nishimoto$^{1,2}$\thanks{Equal last author.} \\
\texttt{nishimoto.shinji.fbs@osaka-u.ac.jp}
\And
Yu Takagi$^{1,2,3}$\footnotemark[1] \\
\texttt{takagi.yu@nitech.ac.jp}
\\[0.5em]
$^{1}$University of Osaka, Japan \\
$^{2}$National Institute of Information and Communications Technology, Japan \\
$^{3}$Nagoya Institute of Technology, Japan
}

\iclrfinalcopy 
\begin{document}

\maketitle

\begin{figure}[ht] 
  \includegraphics[width=\textwidth]{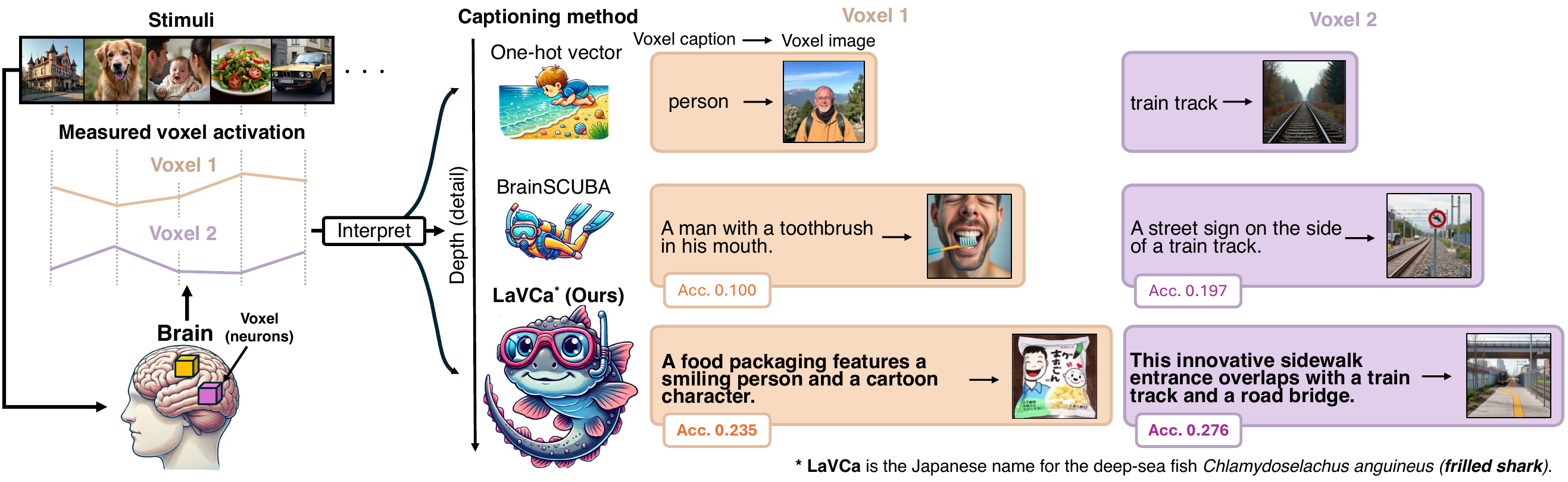}
  \caption{Illustration of our paper. Our proposed method, LaVCa, produces natural language captions that provide a fine-grained description of voxel selectivity (representation) and more accurately capture the characteristics of brain activity in the visual cortex, outperforming conventional approaches.
}
  \label{abstract:abstract_figure}
\end{figure}

\begin{abstract}
Understanding the properties of neural populations (or voxels) in the human brain can advance our comprehension of human perceptual and cognitive processing capabilities and contribute to developing brain-inspired computer models. Recent encoding models using deep neural networks (DNNs) have successfully predicted voxel-wise activity. However, interpreting the properties that explain voxel responses remains challenging because of the black-box nature of DNNs. As a solution, we propose LLM-assisted Visual Cortex Captioning (LaVCa), a data-driven approach that leverages large language models (LLMs) to generate natural-language captions for images to which voxels are selective. By applying LaVCa for image-evoked brain activity, we demonstrate that LaVCa generates captions that describe voxel selectivity more accurately than the previous approaches. The captions generated by LaVCa quantitatively capture more detailed properties than the existing method at both the inter-voxel and intra-voxel levels. Furthermore, we find richer representational content within cortical regions that prior neuroimaging studies have deemed selective for simpler categories. These findings offer profound insights into human visual representations by assigning detailed captions throughout the visual cortex while highlighting the potential of LLM-based methods in understanding brain representations.
\end{abstract}

\section{Introduction}
A primary goal of computer vision is to build systems capable of processing and understanding the complex visual world in a manner akin to human perception. Studying how the human brain—with its advanced visual functions—forms its visual representations deepens our understanding of the brain’s visual network and holds promise for developing next-generation computer vision models.

Over the past decade, \emph{encoding models} have become the standard tool for this endeavour \citep{kay2008identifying, nishimoto2011reconstructing, naselaris2011encoding}.  Early work employed handcrafted, low-level filters or one-hot semantic labels, yielding interpretable—but coarse—descriptions of voxel-level (the spatial measurement unit of fMRI) selectivity. Modern approaches substitute deep neural-network (DNN) features, which dramatically raise prediction accuracy \citep{gucclu2015deep, schrimpf2021neural, takagi2023high}.  Yet the very richness that makes DNNs powerful also renders them opaque: it remains difficult to explain why a given voxel activates, especially at the single-voxel level where group-averaged semantic axes \citep{huth2016natural, lescroart2019human} are too blunt.

In this study, we address the difficulty of voxel-level interpretation with a new method called LLM-assisted Visual Cortex Captioning (\textbf{LaVCa}), which generates data-driven captions for individual voxels (Figure \ref{abstract:abstract_figure}). LaVCa proceeds in four steps: (1) building voxel-wise encoding models for brain activity evoked by images, (2) identifying the optimal images for each voxel’s encoding model using an augmented image dataset, (3) generating captions for these optimal images, and (4) creating concise summaries from those captions. By leveraging large language models (LLMs) with access to a vast, open-ended vocabulary, LaVCa generates diverse inter-voxel captions. Moreover, generating captions from multiple keywords enables us to capture diverse intra-voxel properties.

\begin{wrapfigure}{r}{0.45\textwidth}
    \centering
    \includegraphics[width=0.95\linewidth]{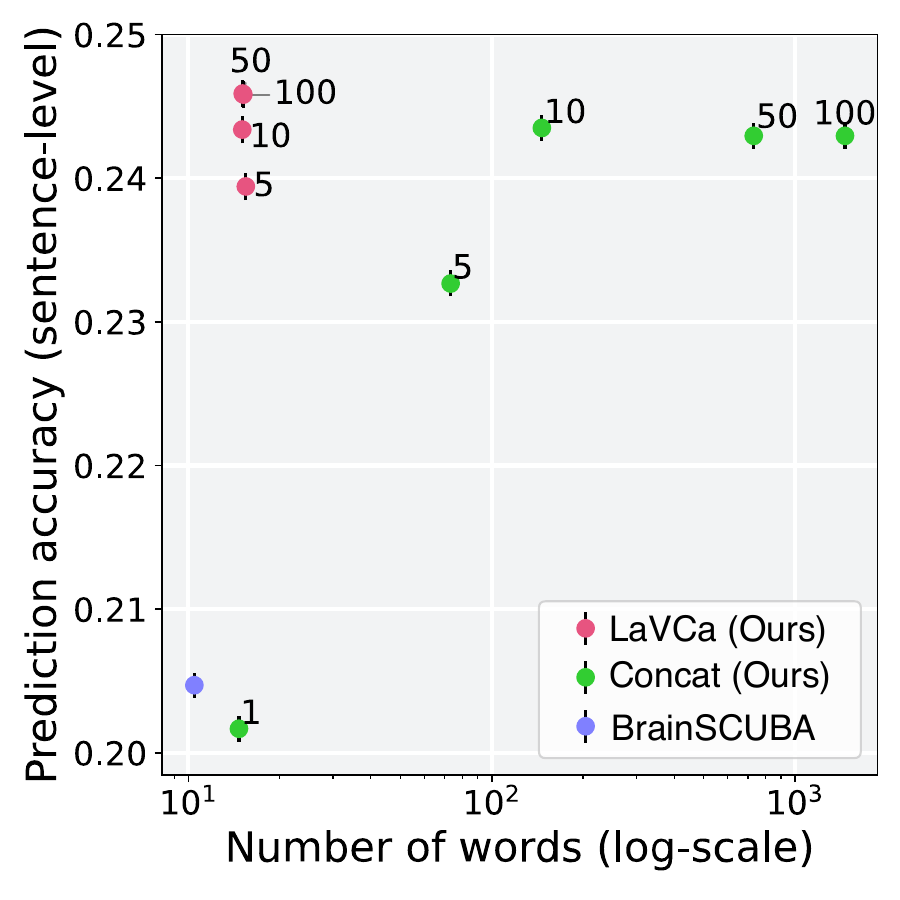}
    \caption{The relationship between brain activity prediction accuracy and voxel caption length (number of words) for a single subject (subj01). Numbers next to each point denote the number of optimal image captions employed by LaVCa for summarization and by Concat for direct concatenation.}
    \label{results:xaxis_num_words}
\end{wrapfigure}

Our contributions are as follows:

\begin{enumerate} 
\item We propose LaVCa, which leverages LLMs to generate natural language captions of voxel-level visual selectivity. By adopting a multi-stage design that decomposes the captioning process into interpretable steps, LaVCa enhances interpretability compared to prior work while preserving descriptive richness.

\item We demonstrate that LaVCa produces more accurate captions than the earlier method BrainSCUBA~\citep{luo2023brainscuba} and better characterizes voxel-wise visual selectivity through brain activity prediction.

\item We also demonstrate that LaVCa can generate highly interpretable and accurate captions without sacrificing information from the optimal images (Figure \ref{results:xaxis_num_words}).

\item The captions generated by LaVCa quantitatively capture more detailed properties than the existing method at both the inter-voxel and intra-voxel levels. 

\item More detailed analysis of the voxel-specific properties generated by LaVCa reveals richer representational content within ROIs that prior neuroimaging studies have deemed selective for simpler categories.

\end{enumerate}


\section{Related Work}

Two complementary approaches frame modern fMRI research: \emph{encoding} and \emph{decoding} models \citep{naselaris2011encoding}.
Encoding models aggregate activity across many \emph{stimuli} for each voxel to pinpoint the features that best explain its responses.  
Decoding models reverse the mapping, pooling activity across many \emph{voxels} to reconstruct a participant’s moment-to-moment percepts—ranging from continuous speech \citep{tang2023semantic} to images \citep{takagi2023high}. Relatedly, recent brain-to-text decoding studies align brain activity with models to generate natural-language descriptions directly from neural signals \citep{chen2025bridging,chen2025mindgpt}. 
Although both lines of work exploit powerful deep models, they address distinct questions: decoders ask “\emph{What was the observer perceiving?},” whereas encoders ask “\emph{What information does this voxel represent?}.”  
The present study targets the encoding side, judging captions by how accurately they predict voxel responses rather than how faithfully they reproduce the original stimulus.

Early encoding work used handcrafted low-level filters or one-hot semantic labels, enabling straightforward but coarse voxel interpretation \citep{kay2008identifying,nishimoto2011reconstructing,naselaris2011encoding,huth2012continuous}. Swapping these features for deep-neural-network (DNN) embeddings dramatically improves prediction accuracy \citep{gucclu2015deep,schrimpf2021neural,takagi2023high,antonello2024scaling}, yet the high-dimensional representations make individual voxels hard to explain. Population-level remedies project many voxels onto a few semantic axes \citep{huth2016natural,lescroart2019human,nakagi2024unveiling}, but sacrifice single-voxel nuance.

To obtain finer, voxel-specific explanations, data-driven text-generation approaches such as BrainSCUBA \citep{luo2023brainscuba} and SASC \citep{singh2023explainingblackboxtext} have been proposed. BrainSCUBA is an end-to-end method that uses an existing image captioning model to produce voxel-wise captions for the visual cortex, whereas SASC uses an LLM to merge multiple short phrases—those with the highest predicted voxel activations—into a single, data-driven caption, thus describing the semantic properties of voxels. However, their reliance on a single captioning model (BrainSCUBA) or on very short n-gram phrases (SASC) limits lexical richness and adaptability.

By (i) decoupling image selection from caption generation, (ii) using LLM-based keyword extraction followed by lightweight sentence composition (iii) allowing any vision-language backbone, and (iv) working with any LLM that has strong language skills without task‑specific fine‑tuning, \textbf{LaVCa retains the high predictive power of brain activity while yielding richer and more controllable voxel-level descriptions than prior work}.

\section{Methods}
\label{main:methods}
\begin{figure}[t]
  \centering
  \includegraphics[width=\textwidth]{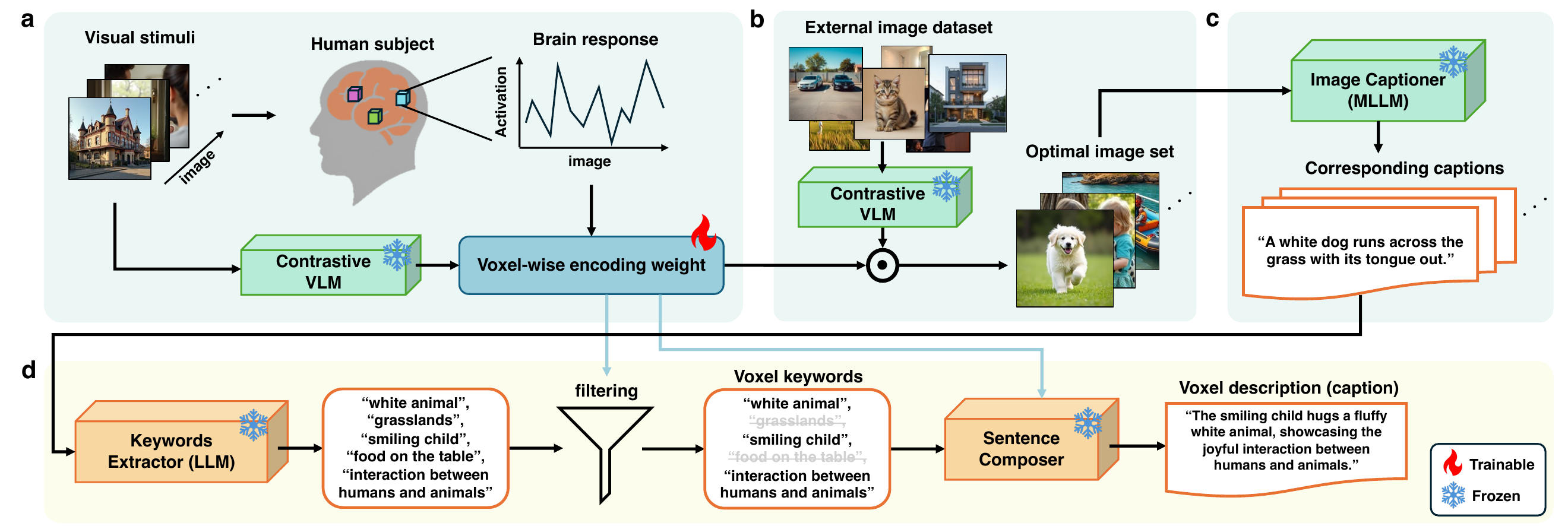}
  \caption{Architecture of LaVCa. \textbf{a} We construct a voxel-wise encoding model for a human subject’s brain activity data (measured using fMRI) while viewing images, using latent representations from a contrastive vision–language model (VLM). The encoding weight is obtained through ridge regression. \textbf{b} We identify the optimal images for a given voxel by calculating the inner product between the contrastive VLM embeddings of external image datasets and the voxel’s trained encoding weight, selecting the top-N images (the ``optimal image set'') that produce the highest predicted activation. \textbf{c} Next, we use a multimodal large language model (MLLM) to generate captions for each optimal image set, allowing an LLM to interpret them. \textbf{d} Finally, we prompt an LLM to extract keywords from the captions, filter these keywords, and feed them into a ``Sentence Composer,'' producing a concise voxel caption.}
  \label{methods:fig1}
\end{figure}

\subsection{fMRI dataset}
This study uses the Natural Scenes Dataset (NSD)~\citep{allen2022massive} following the same experimental conditions as in BrainSCUBA. The NSD consists of data collected over 30 to 40 sessions using a 7 Tesla fMRI scanner, with each participant viewing 10,000 images, repeated three times. We analyze data from the four participants (Subject~01, Subject~02, Subject~05, and Subject~07) who completed all imaging sessions.
The images and captions used in NSD are drawn from MS~COCO and resized to $224 \times 224$ pixels to align with the input requirements of the vision models used. We average the brain activity data for each subject across repeated trials of the same image to improve the signal-to-noise ratio. Up to 9,000 images per subject are used as training data, and the remaining 1,000 images are reserved for testing. We use the preprocessed scans with a resolution of $1.8\,\mathrm{mm}$ provided by NSD for the functional data. We use single-trial beta weights estimated via a generalized linear model within ROIs. Moreover, we standardize the response of each voxel to have a mean of zero and a variance of one within each session. We use the ROIs provided by NSD, which include early and higher-level (ventral) visual areas and face, place, body, and word-selective regions.


\subsection{LLM-assisted Visual Cortex Captioning (LaVCa)}
We propose a method, \textbf{LaVCa (LLM-assisted Visual Cortex Captioning)}, to automatically generate data-driven natural language captions that characterize each voxel’s selectivity in the visual cortex. LaVCa consists of four stages (Figure \ref{methods:fig1}): 

\begin{enumerate}
    \item Construct voxel-wise encoding models for each subject while they view natural images.
    \item Identify the optimal image set by finding the top-$N$ images that most strongly activate each voxel (according to the trained encoding models).
    \item Generate captions for these optimal images using a multimodal large language model (MLLM) for summarization by an LLM in the next step.
    \item Derive concise voxel captions by extracting and filtering keywords from the image captions, then feeding these keywords into a ``Sentence Composer.''
\end{enumerate}

We describe the core pipeline here; ablations are detailed in Appendix~\ref{appendix:ablation}.

\subsubsection{Encoding Model Construction}
First, we construct voxel-wise encoding models to predict each voxel’s activity in response to natural images (Figure~\ref{methods:fig1}a). 
To obtain high-level feature representations of visual stimuli that can be linked to neural responses, 
we use embeddings from a contrastive vision–language model (VLM; e.g., CLIP~\citep{radford2021learning}). 
Specifically, for comparability with BrainSCUBA, we adopt the projection layer embedding of CLIP’s vision branch and use the same pretrained checkpoint as reported in that work (see Appendix~\ref{appendix:pretrained_ckpts}). 
We also re-implemented BrainSCUBA in-house; note that our implementation differs in the dataset used for the projection step and in the training approach for the encoding model (Appendix~\ref{appendix:brainscuba} for details).

For each image stimulus \(i\), we extract its CLIP-Vision projection-layer embedding
\(\mathbf{x}_{i}\in\mathbb{R}^{d}\), L2-normalize it to unit norm, and pair it with the measured responses across all voxels
\(\mathbf{y}_{i}\in\mathbb{R}^{v}\).
The encoding model assumes a linear relationship
\[
\mathbf{y}_{i}= \mathbf{W}\,\mathbf{x}_{i}+\boldsymbol{\varepsilon}_{i},
\]
where \(\mathbf{W}\in\mathbb{R}^{v\times d}\) represents the voxel-wise encoding parameters and \(\boldsymbol{\varepsilon}_{i}\) captures residual noise.
We estimate \(\mathbf{W}\) using ridge regression on the NSD training set.

\subsubsection{Exploration of Optimal Image Sets for Voxels}
Next, we identify the optimal image set for each voxel (Figure \ref{methods:fig1}b). We compute the inner product between the voxel’s encoding weight and CLIP-Vision latent representations from a large-scale external dataset (distinct from NSD) to obtain predicted voxel responses for each image. We then select the top-$N$ images that generate the highest predicted activation. This process is equivalent to calculating the predicted responses of each voxel for every image. This study uses approximately 1.7 million images from OpenImages-v6~\citep{OpenImages} 


\subsubsection{Captioning Optimal Image Sets with MLLM}
To enable an LLM to interpret each voxel’s optimal image set, we first generate captions for these image sets using an MLLM. We use MiniCPM-V~\citep{yao2024minicpmv} with the prompt \textit{``Describe the image briefly.''} For our accuracy evaluation, we also form a simple baseline by concatenating the top-$N$ captions from the optimal image set.

\subsubsection{Generating Voxel Captions}
Finally, we generate interpretable voxel captions from the image captions. First, we use an LLM to extract common keywords across the captions within each voxel’s optimal image set (Figure~\ref{methods:fig1}d). Following the in-context learning prompt approach from~\citep{dunlap2024describing}, we extract multiple keywords from the caption sets using an LLM (\ref{appendix:prompt} for the prompt). We use \textit{gpt-4o} (gpt-4o-2024–08–06 in the OpenAI API) as the LLM. 
To remove irrelevant or noisy keywords, we compute the cosine similarity between each keyword’s embedding from CLIP' text branch (prompted as \textit{``A photo of \{keyword\}.''}) and the encoding weight for that voxel, then apply a softmax threshold to retain only sufficiently relevant keywords. Hereafter, we refer to CLIP's text branch as ``CLIP-Text''.
Next, we transform these filtered keywords into a sentence-level caption using the ``Sentence Composer'' from MeaCap~\citep{zeng2024meacap}, initially designed to generate image captions from keyword sets. 
MeaCap can generate a caption by inputting the target image’s keywords into the Sentence Composer while referencing similarities to the image features.
In this study, we replace image features with encoding weights so that the model composes a coherent sentence from the voxel-specific keywords (for details, see Section~\ref{appendix:meacap}).

\subsection{Caption Evaluation}

\subsubsection{Brain Activity Prediction at Sentence Level}
A voxel caption that truly reflects a voxel’s selectivity should be more similar to the caption of an NSD image that strongly activates that voxel, and less similar to captions of images that do not. We therefore predict voxel-wise brain activity from sentence similarity to evaluate how accurately each caption captures voxel selectivity (Figure \ref{appendix:voxel_pred_method}a). Importantly, this procedure differs conceptually from decoding, which predicts a caption for every stimulus. Following~\citep{singh2023explainingblackboxtext}, we:
\begin{enumerate}
    \item Use a pretrained Sentence-BERT to compute text embeddings for each voxel caption and each NSD image caption.
    \item Compute the cosine similarity between the voxel caption embedding and each NSD image caption embedding.
    \item Treat this similarity value as the predicted activity for that voxel on that image.
\end{enumerate}
For each voxel \(v\), we then calculate the Spearman's rank correlation between the vector of predicted sentence-level similarities and the measured activity; this correlation coefficient is regarded as the prediction accuracy for that voxel.

For statistical significance, we use a permutation test to assess voxel-wise prediction accuracy. Multiple comparisons are corrected using the Benjamini–Hochberg false discovery rate procedure ($\alpha = 0.05$). Detailed procedures are provided in Appendix~\ref{appendix:stats}.

\subsubsection{Brain Activity Prediction at Image Level}
Because sentence-based evaluation can be influenced by non-visual linguistic features (e.g., sentence length, clarity of phrasing, or stylistic variation), we also assess voxel selectivity using \textit{image} similarity (Figure \ref{appendix:voxel_pred_method}b). We use FLUX.1-schnell to create a \textit{voxel image} and then compute vision embeddings (via CLIP-Vision) for both the generated voxel image and each NSD trial image. We obtain an image-level metric of predicted brain activity by comparing these embeddings, focusing purely on visual content. Crucially, this procedure is not image reconstruction in the decoding sense; it characterises voxel selectivity through images rather than attempting to recreate the stimuli themselves.
\section{Results}
\begin{figure}[t]
  \centering
  \includegraphics[width=0.95\textwidth]{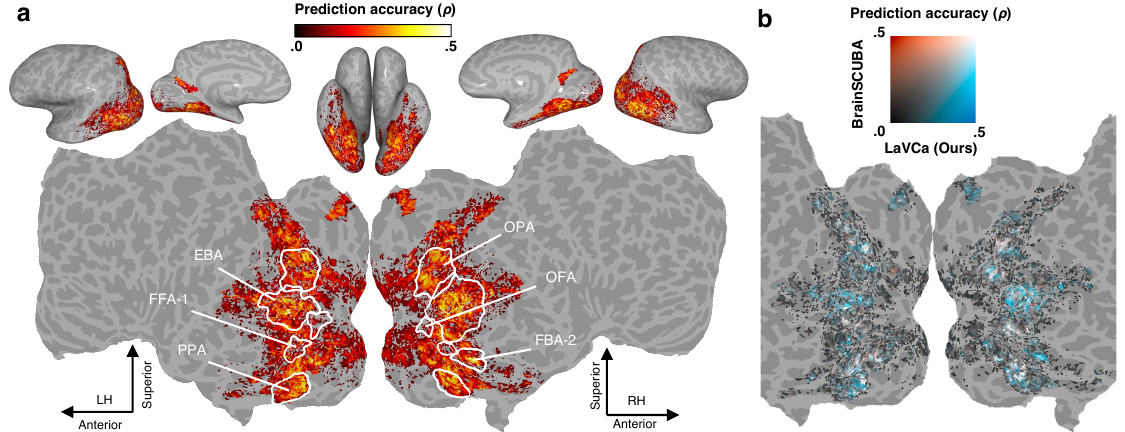}
  \caption{Mapping of brain activity prediction accuracy (subj01). \textbf{a} The sentence-level prediction performance is projected onto inflated cortical surfaces
  (top: lateral, medial, and dorsal views) and flattened cortical surfaces (bottom, with the occipital areas at the center) for both hemispheres. Voxels with significant prediction performance are color-coded (all colored voxels $P<0.05$, FDR corrected). The white outlines indicate the ROIs that are among the top two in terms of the total voxel count across subjects for each semantic category—Body (Extra Striate Body Area; EBA, and Fusiform Body Area; FBA-2), Face (Fusiform Face Area; FFA-1, and Occipital Face Area; OFA), and Places (Parahippocampal Place Area; PPA, and Occipital Place Area; OPA). Word areas are shown in Figure \ref{appendix:sentence_cc_flatmap}. \textbf{b} A comparison of sentence-level prediction performance between our method, LaVCa, and the existing method, BrainSCUBA on the flattened cortical surface. }
   \label{results:accuracy_mapping}
\end{figure}
\subsection{Voxel Activity Prediction}
We examine whether LaVCa can generate concise and interpretable voxel captions without losing critical information in each voxel’s optimal image set. We compare two approaches from the perspective of interpretability by varying the number of optimal images used by LaVCa (Top-$N$) and by simply concatenating the captions of the optimal images (Concat-$N$). Figure~\ref{results:xaxis_num_words} plots prediction accuracy against the average caption length on the horizontal axis, highlighting the trade-off between accuracy and interpretability. Concat-$N$ achieves better accuracy as $N$ increases (up to $N=10$) but at the cost of a much longer caption, which can reduce interpretability. In contrast, LaVCa merges information across the optimal image set into a concise summary, retaining interpretability even as $N$ grows and reaching accuracy comparable to Concat-$N$ (see Figure \ref{appendix:xaxis_words_all_sub}b). Results for all participants are provided in Figure~\ref{appendix:xaxis_words_all_sub}a.

\definecolor{lavca_bg}{HTML}{B5EAD7} 

Next, we determine whether the generated captions accurately capture the properties of voxels in the visual cortex. To this end, we map sentence-level prediction accuracy onto both inflated and flattened cortical surfaces (Figure \ref{results:accuracy_mapping}a). These maps illustrate that LaVCa captions significantly predict voxel activity throughout the visual cortex ($P<0.05$, FDR-corrected). Results for all subjects at the sentence and image levels are presented in Figures \ref{appendix:sentence_cc_flatmap} and \ref{appendix:image_cc_flatmap}.

Finally, we compare two configurations of LaVCa—its default five-keyword version with the Sentence Composer and a simplified single-keyword variant without the Sentence Composer—against the existing method BrainSCUBA and a shuffled variant (LaVCa captions shuffled across voxels) at both the sentence and image levels, focusing on the top 5,000 voxels with the highest accuracy on the training data (Table \ref{table:prediction_accuracy_combined}). Our proposed method, LaVCa, outperforms BrainSCUBA and the single-keyword variant ($P<0.05$, paired $t$-test). This finding suggests that using multiple keywords and composing them into a coherent sentence provides a more accurate explanation of voxel selectivity. Importantly, LaVCa outperforms BrainSCUBA at the \textit{image}-level, indicating that the improvement is not merely due to better handling of non-visual linguistic features (e.g., sentence length or phrasing), but reflects a genuinely enhanced characterization of visual selectivity. 
Furthermore, LaVCa achieves far higher accuracy than the shuffled condition. Results for the top 1,000, 3,000, and 10,000 voxels appear in Table \ref{appendix:TopN_sentence_acc_comparison} and \ref{appendix:TopN_image_acc_comparison}. After visualizing sentence-level prediction accuracy across the cortex, we find that LaVCa exceeds BrainSCUBA’s performance throughout the visual cortex (Figure \ref{results:accuracy_mapping}b). See Figure \ref{appendix:sentence_cc_flatmap} for the results of all subjects.

\begin{table}[t]
\centering
\caption{Comparison of brain activity prediction accuracy at the sentence and image levels. For each subject, the mean and standard deviation of accuracy on the test data are displayed for the top 5,000 voxels with the highest accuracy on the train data.}
\vspace{1.0em}
\scriptsize
\setlength{\tabcolsep}{5pt}
\renewcommand{\arraystretch}{0.9}
\begin{tabular}{lcccccc}
\toprule
\multicolumn{7}{c}{\textbf{Sentence level}}\\
\cmidrule(lr){4-7}
\textbf{Model} & \#\,keywords & Sentence Composer & subj01 & subj02 & subj05 & subj07\\
\midrule
Shuffled       & -- & -- & 0.007 ± 0.199 & 0.058 ± 0.223 & 0.068 ± 0.243 & 0.009 ± 0.175\\
BrainSCUBA     & --  & -- & 0.207 ± 0.062 & 0.251 ± 0.071 & 0.264 ± 0.084 & 0.182 ± 0.065\\
LaVCa (Ours) & 1  & \ding{55} & 0.205±0.068 & 0.250±0.075 & 0.272±0.086 & 0.186±0.072\\
\rowcolor{lavca_bg}
\textbf{LaVCa (Ours)}
               & 5  & \ding{51}  & \textbf{0.246 ± 0.066} & \textbf{0.287 ± 0.075}
               & \textbf{0.306 ± 0.084} & \textbf{0.218 ± 0.073}\\
\midrule
\multicolumn{7}{c}{\textbf{Image level}}\\
\cmidrule(lr){4-7}
\textbf{Model} & \#\,keywords & Sentence Composer & subj01 & subj02 & subj05 & subj07\\
\midrule
Shuffled       & -- & -- & 0.017 ± 0.163 & 0.052 ± 0.185 & 0.066 ± 0.204 & 0.009 ± 0.149\\
BrainSCUBA     & --  & -- & 0.188 ± 0.067 & 0.226 ± 0.070 & 0.250 ± 0.078 & 0.169 ± 0.069\\
LaVCa (Ours) & 1  & \ding{55} & 0.182 ± 0.063 & 0.221 ± 0.066 & 0.252 ± 0.077 & 0.158 ± 0.064\\
\rowcolor{lavca_bg}
\textbf{LaVCa (Ours)}
               & 5  & \ding{51}  & \textbf{0.213 ± 0.072} & \textbf{0.250 ± 0.070}
               & \textbf{0.273 ± 0.079} & \textbf{0.187 ± 0.073}\\
\bottomrule
\end{tabular}
\label{table:prediction_accuracy_combined}
\end{table}

\subsection{Lexical and Semantic Diversity Analysis}
We next assess how effectively LaVCa captions capture both lexical and semantic diversity across voxels, focusing first on \textit{inter-voxel} diversity (Table \ref{table:diversity_analysis}, left). For this quantitative evaluation, we use three metrics: (1) the total vocabulary size (excluding stop-words) across all voxel captions (Lexical); (2) the average variance across each dimension of the CLIP-Text embedding computed on all voxel captions (Semantic); and (3) the number of principal components (PCs) required to capture 90\% of the variance of CLIP-Text embedding across captions in a principal component analysis (PCA; Semantic). 

First, we evaluate the diversity of LaVCa captions compared with the existing method, BrainSCUBA. When averaged across subjects, LaVCa markedly outperforms BrainSCUBA in both lexical (16,922 vs. 3,193 in vocab. size) and semantic (0.0642 vs. 0.0588 in variance of embeddings; 219 vs. 127 in PCs required for 90\% variance explained) diversity. These findings confirm that our open-ended LLM–based approach can produce richer word usage and more meaningful captions across inter-voxel comparisons. 

We evaluate the diversity of LaVCa captions compared with more detailed captions. BrainSCUBA leverages ClipCap \citep{mokady2021clipcap}, a model that produces relatively simple image captions. We use the top-1 captions generated by the MLLM on the optimal image sets (equivalent to the case where $N=1$ in Concat-$N$) to compare the diversity of LaVCa with more detailed captions. When averaged across subjects, Top-$1$ (13,959 vocab. size, 0.0638 avg. variance, 210 PCs) exhibits both a vocabulary range and semantic diversity close to LaVCa. However, LaVCa achieves a higher prediction accuracy (0.264 vs. 0.224), indicating that LaVCa can preserve robust brain activity prediction performance while enhancing the diversity of generated captions.

Next, we evaluate diversity from an \textit{intra-voxel} perspective by comparing captions generated by three models in both lexical and semantic dimensions (Table \ref{table:diversity_analysis}, right). We use three metrics: (1) the vocabulary size of each voxel’s caption (Lexical), (2) the average sentence length in each voxel’s caption (Lexical), and (3) the average variance across all dimensions of Word2Vec embeddings of each caption’s words (excluding stop-words) (Semantic). When averaged across subjects, LaVCa markedly outperforms BrainSCUBA in both lexical (11.4 vs. 6.09 in vocab. size) and semantic (11.9 vs. 6.19 in avg. length; 0.0199 vs. 0.0160 in variance of semantic embeddings) diversity. This improvement suggests that LaVCa more precisely captures the fine-grained intra-voxel characteristics. 

For examples of voxel captions and images from various OFA and PPA voxels—along with their corresponding quantitative metrics—compared across three models (LaVCa, BrainSCUBA, Top-1), see Figures \ref{appendix:OFA_example_1}, \ref{appendix:OFA_example_2}, \ref{appendix:PPA_example_1}, and \ref{appendix:PPA_example_2}.

\subsection{ROI-level Diversity Analysis}
Our results thus far demonstrate that LaVCa produces more accurate voxel captions than BrainSCUBA and better captures both inter- and intra-voxel diversity. We next ask whether LaVCa can reveal richer representational content inside ROIs that earlier neuroimaging studies have largely described as selective for simpler categories—for example, faces in the OFA or places in the PPA \citep{gauthier2000fusiform, haxby2000distributed, epstein1998cortical}. We conduct a qualitative and quantitative evaluation using LaVCa’s captions and generated images to analyze diversity that exists beyond the known selectivity in the ROI.

\begin{figure}[t]
    \centering
    \includegraphics[width=0.91\textwidth]{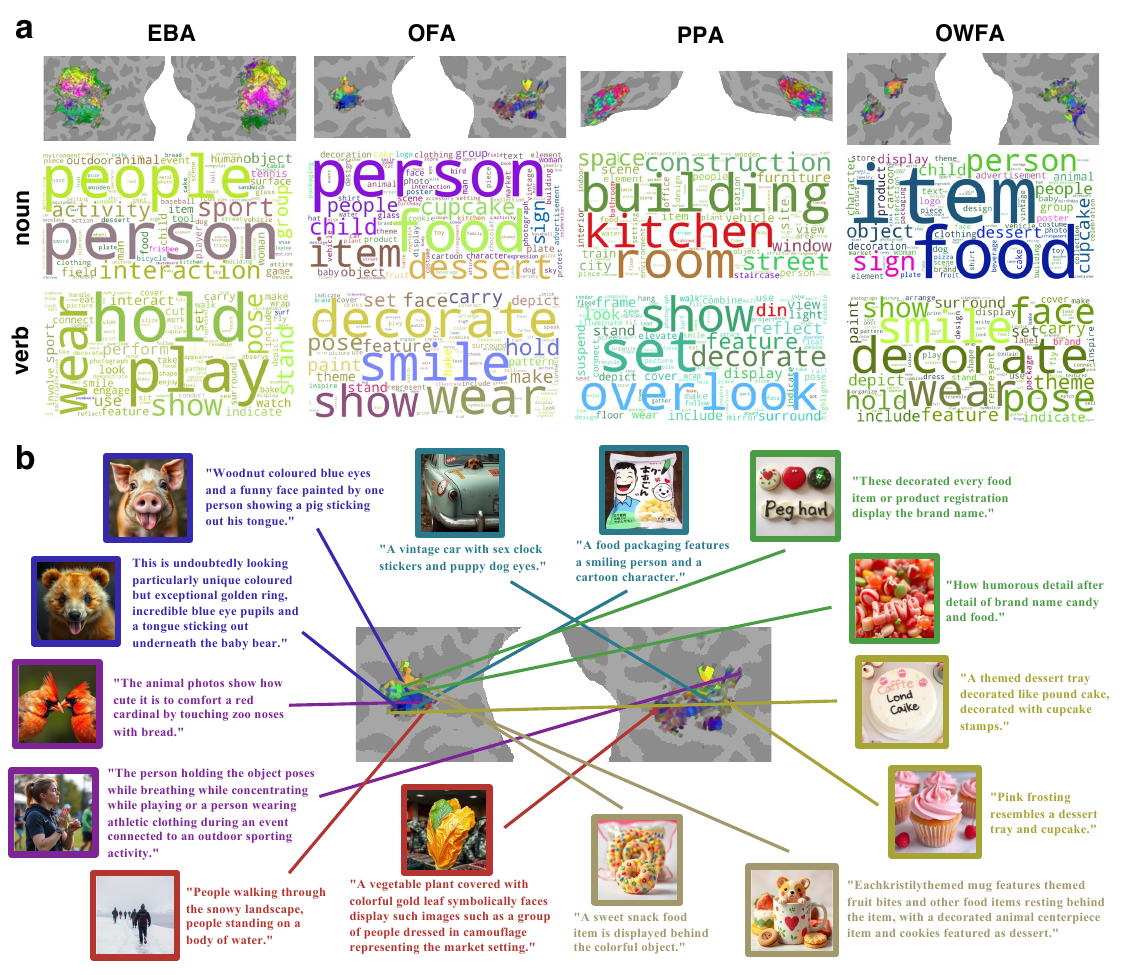}
    \caption{Interpretation of LaVCa captions (subj02). 
    \textbf{a} UMAP projection of caption text across four ROIs (EBA, OFA, PPA, OWFA), visualized on a flatmap (top). 
    A word cloud of the 100 most frequent \textbf{nouns} in these captions (middle), colored by location in the UMAP space. 
    A word cloud of the 100 most frequent \textbf{verbs} (bottom). 
    \textbf{b} Visualization of the top two captions (by accuracy) for eight clusters on the flatmap in OFA. 
    The images generated for each caption appear to the left or above the text. Voxels are connected to their corresponding captions and images by lines. 
    The color of each caption and image border reflects the average UMAP color of all voxels in the cluster.}
    \label{results:OFA_caption_vis}
\end{figure}

\paragraph{Qualitative Assessment.}
We explore the semantic diversity of LaVCa captions across four ROIs (EBA, OFA, PPA, OWFA) by applying UMAP to their CLIP-Text embeddings and visualizing the resulting distributions on a flatmap (Figure~\ref{results:OFA_caption_vis}a, top). In each ROI, we observe a broad spectrum of UMAP colors, indicating multiple meaningful clusters within regions known for distinct category-selective responses. The presence of this broad spectrum is consistent across participants (Figure~\ref{appendix:EBA_OFA_WordCloud}, \ref{appendix:PPA_OWFA_WordCloud}).

Across ROIs, we observe diverse nouns and verbs that not only align with prior selectivity profiles but also reveal richer, voxel-level selectivity for object categories and actions. Both EBA and OFA frequently include common nouns such as people'' and person,'' and the verb distributions highlight ROI-specific action tendencies: EBA is enriched for body-related actions (e.g., hold''), whereas OFA is enriched for face-related actions (e.g., smile''). These patterns are consistent across participants (Figure~\ref{appendix:EBA_OFA_WordCloud}, \ref{appendix:PPA_OWFA_WordCloud}).

Finally, to highlight how each caption and its corresponding voxel image relate to specific colors in semantic space, we project them onto a flatmap (Figure \ref{results:OFA_caption_vis}b). We divide the samples into eight clusters by labeling each of the three UMAP dimensions as ``High'' ($\geq2/3$) or ``Low'' ($\leq1/3$). From each cluster, we pick the two voxels with the highest prediction accuracy (or one if only one qualifies, or none if none qualify) and illustrate their captions and generated images.

In OFA, some captions are related to faces (e.g., ``face,'' ``person,'' ``animal''), while particular voxels encoded more fine-grained features such as ``eye,'' ``tongue,'' or ``smiling,'' and other voxels encoded information like ``animal,'' ``bear,'' or ``cardinal.'' Thus, even within this ROI, there appears to be substantial functional differentiation among inter-voxel that extends beyond a generic ``face'' category.

Moreover, we observe \textit{intra-voxel} diversity, where a single caption incorporates multiple ideas (e.g., \textit{``A food packaging features a smiling person and a cartoon character''}), suggesting that individual voxels can simultaneously encode several distinct concepts. 
These findings highlight the fine-grained functional specialization across inter-voxel within the ROI and the diverse nature of intra-voxel encoding beyond singular concepts.

The results for all participants, visualizing the top two captions for each cluster directly in the UMAP space, can be found in Figures \ref{appendix:OFA_caps_umap_subj01-02}, \ref{appendix:OFA_caps_umap_subj05-07}, \ref{appendix:PPA_caps_umap_subj01-02}, and \ref{appendix:PPA_caps_umap_subj05-07}.

\paragraph{Quantitative Assessment.}
We next determine how many distinct captions appear in each ROI by comparing the sentence-level prediction accuracy of each ROI when captions are maintained in their original form versus shuffled within the ROI (Table~\ref{results:ROI_caption_shuffle}). For each category (body, face, place, and word area), we select two ROIs with the largest total voxel count across all subjects, resulting in eight ROIs in total. In all ROIs, shuffling reduces prediction accuracy significantly. For example, in the OFA, accuracy drops from 0.0945 (Original) to 0.0280 (Shuffled), a 3.3-fold decrease; in the PPA, accuracy falls from 0.213 (Original) to 0.151 (Shuffled), a 1.4-fold decrease. Thus, even in regions traditionally linked to particular concepts, voxels exhibit a range of distinct selectivities. Furthermore, the average caption similarity between the same ROIs of different subjects is relatively high at 0.227, compared to 0.171 between different ROIs of different subjects, indicating that such diversity is reproducible across subjects (Figure \ref{results:inter-roi_inter-subj}).

Next, we quantify how many different semantic concepts a single voxel can encode (i.e., its degree of multi-concept selectivity). We perform the following analysis: (1) extract every unique noun from all voxel captions within the ROI; (2) obtain CLIP-Text embeddings for each noun using the prompt ``A photo of \{word\}.'' and cluster them with \(k\)-means (\(k=6\)); (3) for each voxel, count how many of its nouns fall into different clusters. Across all ROIs, we find that most voxels are associated with multiple clusters, indicating multi-concept selectivity (Table~\ref{table:multi_concept}). Thus, even within ROIs whose vocabulary is relatively coherent, individual voxels can encode several distinct concepts. Furthermore, by aggregating the nouns used in this clustering analysis from all subjects and examining the extent to which each subject’s voxels belong to the subject-shared clusters, we evaluate the cross-subject reproducibility of ROI diversity (Figure \ref{appendix:inter-subject-cluster-anlysis}). In both the OFA and PPA, voxels from all subjects populate the same clusters, suggesting that such diversity is, to some extent, consistent across individuals.

\begin{table}[t]  
\centering
\caption{Average prediction accuracy with standard error across subjects when captions within each ROI are shuffled (Shuffled) versus used as is (Original).}
\vspace{1.0em}
\scriptsize
\setlength{\tabcolsep}{4pt}      
\renewcommand{\arraystretch}{0.9} 
\begin{tabular}{c*{8}{c}}
\toprule
& \multicolumn{2}{c}{\textbf{Body areas}}
& \multicolumn{2}{c}{\textbf{Face areas}}
& \multicolumn{2}{c}{\textbf{Place areas}}
& \multicolumn{2}{c}{\textbf{Word areas}} \\
\cmidrule(lr){2-3}\cmidrule(lr){4-5}\cmidrule(lr){6-7}\cmidrule(lr){8-9}
Model & EBA & FBA-2 & OFA & FFA-1 & OPA & PPA & OWFA & VWFA-1 \\
\midrule
Shuffled & 0.018±0.008 & 0.018±0.005 & 0.028±0.004 & 0.016±0.003
         & 0.116±0.024 & 0.151±0.028 & 0.025±0.005 & 0.034±0.009 \\
\rowcolor{lavca_bg}
\textbf{Original} & \textbf{0.157±0.005} & \textbf{0.125±0.010} & \textbf{0.095±0.009} & \textbf{0.111±0.003}
                  & \textbf{0.200±0.022} & \textbf{0.213±0.027} & \textbf{0.084±0.013} & \textbf{0.158±0.007} \\
\bottomrule
\end{tabular}
\label{results:ROI_caption_shuffle}
\end{table}


\section{Discussion \& Conclusions}
\label{main:discussion_and_conclusions}
In this study, we introduce a novel method called LaVCa, which leverages LLMs to produce data-driven, natural-language descriptions of voxel selectivity in the human visual cortex. The voxel captions generated by LaVCa exhibit higher accuracy and greater semantic diversity than those generated by the existing approach, BrainSCUBA. We attribute this improvement to our mechanism for integrating multiple keywords extracted by advanced LLMs, which enables a more comprehensive capture of the diverse selectivity patterns across voxels. Furthermore, LaVCa uncovers richer representational content within ROIs that earlier neuroimaging studies had characterized as selective for simpler categories. By revealing that even “category-selective” areas such as the OFA and PPA encode a broader spectrum of concepts, our findings challenge long-standing assumptions about functional specialization in the visual cortex. See Sections \ref{appendix:limitation} and \ref{appendix:impact_statement} for the Limitation and Impact Statement.

\section*{Acknowledgment}
Y.T. was supported by JST PRESTO (Grant Numbers JPMJPR23I6 and JPMJCR2574) and JSPS KAKENHI (Grant Numbers JP19H05725 and JP24K02999). S.N. was supported by JSPS KAKENHI JP23H05493 and JP24H00619 and JST JPMJCR24U2.

\section*{Ethics Statement}
This study did not involve the collection of any new neural recording data. Instead, we relied exclusively on the Natural Scenes Dataset (NSD), which is openly accessible to the research community. The dataset can be obtained from \url{https://naturalscenesdataset.org/}
, subject to their terms of use.

We conducted all analyses on this publicly released dataset and did not handle any personally identifiable information. Based on the nature of the data and the scope of our methods, we do not anticipate harmful applications of this work.

\section*{Reproducibility Statement}
We have made substantial efforts to ensure the reproducibility of our results. 
Details of the LaVCa pipeline, including model architecture, training procedure, and evaluation metrics, are described in Section~\ref{main:methods}. 
Additional implementation details, hyperparameter settings, and preprocessing steps for the Natural Scenes Dataset (NSD) are provided in the Appendix. 
Furthermore, the source code and scripts necessary to reproduce our experiments are publicly available at:
\url{https://github.com/suyamat/LaVCa}.

\section*{LLM Usage}
In accordance with the ICLR policy on the use of large language models (LLMs), 
we report that LLMs were employed exclusively for language-related assistance. 
Specifically, we used LLMs to aid in the translation of text into English and to polish the grammar and style of the manuscript. 
All research ideas, experimental design, data analysis, and scientific interpretations were conceived and conducted entirely by the authors. 

The use of LLMs did not contribute to the formulation of research questions, methodology, or conclusions. 
The authors take full responsibility for the final content of the paper, including all text that was assisted by LLM-based tools.

\bibliography{iclr2026_conference}
\bibliographystyle{iclr2026_conference}

\renewcommand\thefigure{A\arabic{figure}}
\renewcommand{\theHfigure}{A\arabic{figure}}

\setcounter{figure}{0}
\renewcommand\thetable{A\arabic{table}}
\renewcommand{\theHtable}{A\arabic{table}}
\setcounter{table}{0}
\renewcommand\theequation{A\arabic{equation}}
\renewcommand{\theHequation}{A\arabic{equation}}

\setcounter{equation}{0}

\appendix
\clearpage
\onecolumn

\section{Appendix}
\label{appendix:main}

\subsection{Full Related Work}
\label{appendix: related_work_1}
\subsubsection{Interpreting the representations of the brain's neurons.}
Encoding models have long been used in neuroscience to interpret neural representations within the brain~\citep{kay2008identifying, nishimoto2011reconstructing, naselaris2011encoding, huth2012continuous}. These studies used interpretable features, such as low-level visual attributes, or high-level semantic features, such as one-hot encoding of words, for straightforward voxel-wise interpretation.

Recent approaches use features derived from DNNs and have demonstrated higher explanatory power for brain activity than those using simpler, more interpretable features~\citep{gucclu2015deep, schrimpf2021neural, takagi2023high, antonello2024scaling}.  However, the interpretability of these DNN-based encoding models remains challenging, leading to the development of methods that condense the entire set of voxels into a small number of universal and interpretable axes~\citep{huth2016natural, lescroart2019human, nakagi2024unveiling}.

Recent approaches propose data-driven methods to describe the properties of individual brain voxels using natural language~\citep{luo2023brainscuba, singh2023explainingblackboxtext} when analyzing brain representations at a finer, voxel-wise level. BrainSCUBA~\citep{luo2023brainscuba} is an end-to-end method that uses an existing image captioning model, which provides voxel-wise captions of the visual cortex in a data-driven manner. BrainSCUBA projects each voxel‘s encoding weight onto the image feature space via dot-product attention, identifies regions of highest similarity, and then uses a text decoder to generate captions describing the images to which the voxel is most selective. This approach provides a data-driven natural-language description of voxel selectivity without additional training. Similarly, SASC~\citep{singh2023explainingblackboxtext} uses fMRI data collected during speech listening~\citep{lebel2023natural} to identify the short phrases that most strongly activate each voxel. It then uses an LLM to combine these short phrases into a single, data-driven caption describing each voxel’s semantic properties. 

Our proposed method also generates data-driven voxel captions but differs in several ways. First, BrainSCUBA is constrained to pre-existing, end-to-end image captioning models. In contrast, our approach divides the task into (i) identifying an optimal set of images and (ii) converting these images into a caption, allowing us to use any vision model aligned with language and any LLM with advanced language capabilities without requiring specialized fine-tuning. Furthermore, although SASC uses an LLM to create voxel captions, it primarily synthesizes short, low-information phrases (e.g., trigrams), producing only simple keyword-based captions. In contrast, our method summarizes more diverse and informative text and then uses these extracted keywords to compose a complete sentence, capturing a richer range of voxel-level properties.

\subsubsection{Interpreting the Representations of Artificial Neurons in DNNs}
Interpreting artificial neurons is a key challenge in understanding how DNNs process information. We can potentially examine human neural representations at a finer granularity by applying the data-driven and highly accurate interpretation methods developed for artificial neurons to analyze human brain voxels.

Numerous studies have aimed to associate artificial neurons with human-interpretable concepts~\citep{bau2017network, mu2020compositional, oikarinen2022clip, kalibhat2023identifying, bykov2024labeling}. These methods link neurons to textual concepts by comparing neuron output feature maps with outputs from segmentation models. However, these approaches are constrained by predefined concept sets or limited to the dataset’s words and phrases. MILAN~\citep{hernandez2021natural} introduced a generative approach, enabling adaptation to different domains and tasks, but it requires annotated data, which poses challenges for scalable applications.

LLMs permit open-ended descriptions of artificial neurons without additional model training~\citep{singh2023explainingblackboxtext, bai2024describe, wu2024and, hoang2024llm}. Analogous to these methods, our study also leverages LLMs to generate open-ended concepts for \textbf{brain} neurons rather than artificial neurons, seeking flexible and diverse interpretations that do not depend on predefined vocabularies.

\subsection{Implementation Details}
\subsubsection{Generating Voxel Captions}
\label{appendix:meacap}

In this study, we leverage the ``Sentence Composer'' proposed in the image captioning model MeaCap~\citep{zeng2024meacap}—referred to as the ``keywords-to-sentence LM'' in the original paper—to generate sentence-level captions from keywords.

\paragraph{Notation}

\begin{itemize}
  \item $K=\{k_1,\dots ,k_m\}$: keyword set extracted by an LLM.  
  \item $W\in\mathbb{R}^d$: voxel-wise encoding weight.  
  \item $\{\tilde{c}_1,\dots ,\tilde{c}_k\}$: top-$k$ captions of the voxel’s optimal images.  
  \item $\phi_T(\cdot)$: CLIP-Text embedding operator.  
  \item $\displaystyle 
        \operatorname{sim}(\mathbf{u},\mathbf{v})=
        \frac{\mathbf{u}^{\top}\mathbf{v}}
             {\lVert\mathbf{u}\rVert\,\lVert\mathbf{v}\rVert}$: cosine similarity.  
\end{itemize}

\paragraph{Iterative Decoding Procedure}

We begin with an \emph{initial draft caption}
\[
  \mathbf{c}^{(0)} = [k_1\ k_2\ \dots\ k_m],
\]
obtained by concatenating the keywords $K$ in their given order (separated by
spaces).  Starting from this seed, CBART iteratively refines the caption
through the following steps:

\begin{enumerate}
  \item \textbf{Action prediction.}  
        For each position $j$ in the current draft $\mathbf{c}^{(t)}$,  
        the encoder assigns one of the actions  
        $\textsc{Copy}$, $\textsc{Replace}$, or $\textsc{Insert}$.

  \item \textbf{Candidate generation.}  
        At positions marked for replacement or insertion, the decoder proposes  
        the top-$n$ lexical candidates  
        $\mathcal{W}_j=\{w_{j,1},\dots ,w_{j,n}\}$  
        ranked by the token likelihood $P_\theta(w\mid\mathbf{c}_{<j}^{(t)})$.

    \begin{itemize}
      \item \emph{Fluency} ($\log P_\theta$) ensures linguistic naturalness.  
      \item \emph{Image relevance} grounds the sentence in visual evidence from the voxel’s optimal images.  
      \item \emph{Voxel relevance} ($\operatorname{sim}(\phi_T(w),W)$) links the caption to the voxel’s representation.  
    \end{itemize}
    
    Each candidate $w\in\mathcal{W}_j$ is scored by
    \begin{equation}
      S(w)=
        \lambda_1 \log P_\theta\!\bigl(w \mid \mathbf{c}^{(t)}_{<j}\bigr)
        + \lambda_2 \frac{1}{k}\sum_{i=1}^{k}
            \operatorname{sim}\!\bigl(\phi_T(w),\phi_T(\tilde{c}_i)\bigr)
        + \lambda_3 \operatorname{sim}\!\bigl(\phi_T(w),W\bigr),
      \label{eq:score}
    \end{equation}
    where we set $(\lambda_1,\lambda_2,\lambda_3) = (0.2,\,0.2,\,1.2)$.

  \item \textbf{Token selection and refinement.}  
        The token with the highest $S(w)$ replaces or is inserted at position $j$,  
        yielding the updated draft $\mathbf{c}^{(t+1)}$.  
        The loop repeats until every position is predicted as \textsc{Copy},  
        producing the final caption $\hat{\mathbf{c}}$.
\end{enumerate}

\paragraph{Rationale}

\begin{itemize}
  \item \emph{Fluency} $\bigl(\log P_\theta\bigr)$ encourages linguistic naturalness.  
  \item \emph{Image relevance} $\bigl(\frac{1}{k}\sum_{i=1}^{k}
        \operatorname{sim}\bigl(\phi_T(w),\phi_T(\tilde{c}_i)\bigr)\bigr)$ grounds the sentence in visual evidence drawn from the voxel’s optimal images.  
  \item \emph{Voxel relevance} $\bigl(\operatorname{sim}\bigl(\phi_T(w),W\bigr)\bigr)$ ties the caption to the voxel’s learned representation.  
\end{itemize}

By jointly optimizing the score in~\eqref{eq:score},  
the method transforms discrete keyword sets into a coherent sentence that is
\emph{linguistically natural}, \emph{visually grounded}, and \emph{specifically aligned} with the voxel’s
weights~$W$.

\subsubsection{Statistical Testing}
\label{appendix:stats}
For each voxel $v$, the observed prediction accuracy is quantified as Spearman’s rank correlation $\rho_v^{\text{obs}}$. 
To realize the null hypothesis of no association, the activity vector is randomly permuted $B$ times, yielding surrogate correlations $\{\rho^{\text{null}}_{v,b}\}_{b=1}^{B}$. 
Pooling across all $N$ voxels produce a global null distribution. 
The one-tailed $p$-value is computed as
\[
p_v = \frac{\#\{\rho^{\text{null}}_{v,b} \geq \rho^{\text{obs}}_v\} + 1}{B \times N + 1}.
\]
We control for multiple comparisons using the Benjamini--Hochberg false-discovery-rate (FDR) procedure ($\alpha = 0.05$); voxels with $q < 0.05$ are declared significant. 
In our experiments, we set $B=1000$.

\subsubsection{Pretrained Checkpoints}
\label{appendix:pretrained_ckpts}
We rely on a variety of pretrained models for different components of our pipeline. 
Most of the models are publicly available checkpoints hosted on Hugging Face, including contrastive vision--language models (CLIP, SigLIP2, FG-CLIP), multimodal LLMs (MiniCPM-Llama3-V2.5, BLIP), LLM (Llama 3.1-70B), sentence similarity models (Sentence-BERT, MPNet), and a text-to-image model (FLUX.1-schnell). 
For keyword extraction, we use gpt-4o, which is not available as a checkpoint but is accessed via the OpenAI API. 
A complete list of all models and their repositories is summarized in Table~\ref{tab:pretrained_ckpts}.

\begin{table}[t]
\centering
\caption{Pretrained checkpoints used in our experiments.}
\label{tab:pretrained_ckpts}
\resizebox{\textwidth}{!}{
\begin{tabular}{lll}
\toprule
Category & Model & Repository (Hugging Face) \\
\midrule
\multirow{3}{*}{\centering Contrastive VLM}
  & CLIP & \texttt{openai/clip-vit-base-patch32} \\
  & SigLIP2 & \texttt{google/siglip2-base-patch16-224} \\
  & FG-CLIP & \texttt{qihoo360/fg-clip-base} \\
\midrule
\multirow{2}{*}{\centering MLLM}
  & MiniCPM-Llama3-V2.5 & \texttt{openbmb/MiniCPM-Llama3-V-2\_5} \\
  & BLIP & \texttt{Salesforce/blip-image-captioning-base} \\
\midrule
\multirow{2}{*}{\centering LLM}
  & gpt-4o & \textit{N/A (OpenAI API)} \\
  & Llama 3.1-70B & \texttt{meta-llama/Llama-3.1-70B-Instruct} \\
\midrule
\multirow{2}{*}{\centering Sentence Similarity Model}
  & Sentence-BERT & \texttt{sentence-transformers/all-MiniLM-L6-v2} \\
  & MPNet & \texttt{sentence-transformers/all-mpnet-base-v2} \\
\midrule
\multirow{1}{*}{\centering Text-to-Image Model}
  & FLUX.1-schnell & \texttt{black-forest-labs/FLUX.1-schnell} \\
\bottomrule
\end{tabular}
}
\end{table}


\subsubsection{BrainSCUBA}
\label{appendix:brainscuba}
At the outset of our project in January 2025, the original BrainSCUBA codebase had not yet been released, so we implemented the method ourselves for this study. In BrainSCUBA, the encoding weights (linear layers) are learned using gradient descent. In our implementation, consistent with our proposed method, we trained the encoding weights using L2-regularized linear regression from the \textit{himalaya} library package\footnote{\href{https://github.com/gallantlab/himalaya}{https://github.com/gallantlab/himalaya}}~\citep{la2022feature}.

Moreover, BrainSCUBA projects each voxel‘s encoding weight into image space using a dataset of 2 million images, combining OpenImages~\citep{OpenImages} and LAION-A v2 (6+ subset)~\citep{schuhmann2022laion}. However, the specific images selected from each dataset are not disclosed. We ensure a fair and dataset-independent comparison by relying solely on the 1.7 million images from OpenImages (the same dataset used by our proposed method, LaVCa). We leverage the training set of the subset that is accompanied by bounding boxes, object segmentations, visual relationships, and localized narratives.

\begin{figure*}[t] 
  \centering
  \includegraphics[width=\textwidth]{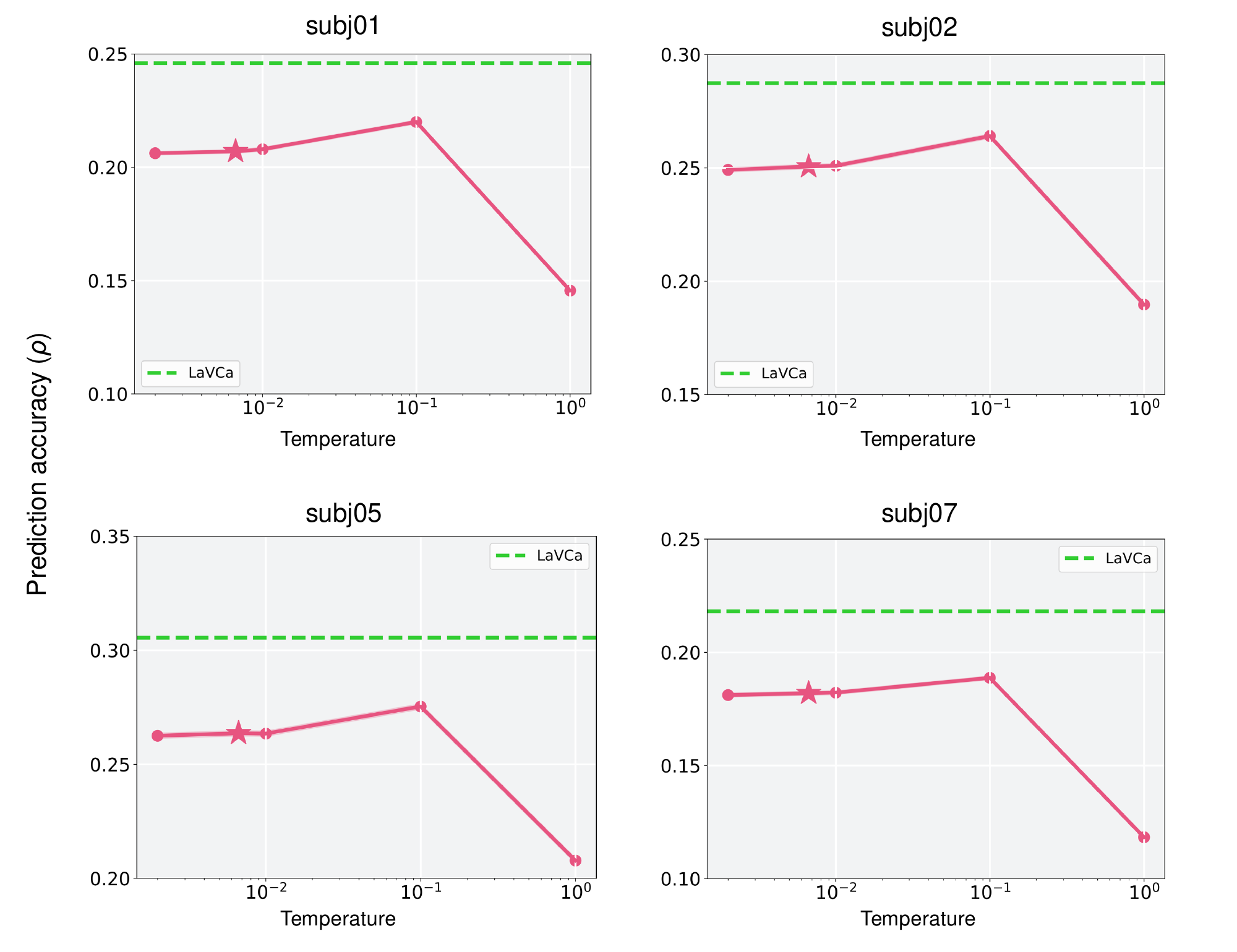}
  \caption{The change in accuracy of BrainSCUBA with respect to temperature. The error bars represent standard error. Moreover, the star markers on the plot indicate the point where the temperature is set to 1/150, as adopted in the original BrainSCUBA paper. The green line represents the average value of LaVCa.
}
  \label{appendix:scuba_temp}
\end{figure*}

For other hyperparameters, we tested temperature values of 1.0, $1/10$, $1/100$, $1/150$ (the value used in the BrainSCUBA paper), and $1/500$ for the softmax projection (Figure \ref{appendix:scuba_temp}). We used beam search with a beam width of 5 to generate the text decoder’s caption as described in the BrainSCUBA paper.

\subsubsection{Used Compute Resources}
\label{appendix:compute_resources}
All experiments are conducted on a single Lambda Labs cloud instance equipped with eight NVIDIA A100-SXM4 GPUs (40 GB each). The host system features a dual-socket AMD EPYC 7542 processor, providing 124 logical CPU cores and 512 GB of DDR4 RAM.

In this setting, voxel-wise Ridge-regression training for one subject finishes in \textasciitilde{}7~s and occupies 5.4~GB of GPU memory.  
Loading the 1.7~M candidate images for optimal-image search, performed on the CPU, takes \textasciitilde{}1,490~s per subject.  
The subsequent per-voxel pipeline---optimal-image search, multimodal-LLM captioning, keyword extraction, and SentenceComposer generation---runs in \textasciitilde{}15~s per voxel with a peak GPU footprint of 5.0~GB.  
Processing 20,000 voxels for a single subject therefore requires \textasciitilde{}83~h end-to-end, which remains practical for offline analyses in systems neuroscience.

\clearpage
\subsection{Ablation Study}
\label{appendix:ablation}
\subsubsection{Encoding Model}
\label{appendix:ablation_encoding}
We conduct an ablation study to evaluate how hyper-parameter choices affect prediction accuracy for the encoding model (Table~\ref{appendix:encoding_compare}).  
Here, “accuracy” denotes the correlation coefficient obtained when predicting brain activity with the encoding model, rather than a voxel-caption–based metric).

\paragraph{Voxel-wise versus shared weights.}
We compare two variants of ridge regression: a \emph{voxel-wise} version that learns an individual weight vector for each voxel and a \emph{shared-weights} version that learns a single weight vector common to all voxels.  
The regularisation coefficient \(\lambda\) is sampled at 25 logarithmically spaced points from \(10^{-4}\) to \(10^{20}\), and the optimal value is selected via five-fold cross-validation.  
This comparison reveals almost no difference in accuracy between the two variants.

\paragraph{Ridge regression versus gradient descent.}
We next contrast linear layers trained with ridge regression against those trained with gradient descent.  
The hyper-parameter settings for gradient-descent training are summarised in Table~\ref{appendix:grad_config}.  
During optimisation, the final 10\,\% of the training data are held out for validation, and early stopping is triggered if validation accuracy fails to improve for five consecutive epochs.  
To ensure a fair comparison, ridge regression is fitted using only the first 90\,\% of the training data, so that both methods see an identical amount of data.  
Under the \emph{shared-weights} setting, ridge regression outperforms gradient descent, indicating that an analytically derived linear solution is more effective for training the encoding model.

\paragraph{Linear versus non-linear models.}
Finally, we compare linear and non-linear architectures.  
For the non-linear networks, the hidden-layer width is set equal to the input dimensionality (512, corresponding to the \textsc{CLIP}-Vision feature size).  
Introducing non-linearities yields higher accuracy than the single-layer linear model trained with gradient descent; however, increasing the depth from two to three layers offers no further benefit.  
Moreover, none of the non-linear networks surpass the performance of ridge regression.  
These findings indicate that, although non-linear models trained with gradient descent can exceed their linear counterparts, they still fall short of the accuracy achieved by ridge regression.

\begin{table}[t]
\centering
\caption{Hyper-parameter settings for gradient-descent training.}
\label{appendix:grad_config}
\begin{tabular}{ll}
\toprule
\textbf{Parameter}                        & \textbf{Value} \\ \midrule
Optimizer                            & AdamW ($\beta_{1}=0.9,\;\beta_{2}=0.999$) \\
Batch size                           & 64 \\
Initial learning rate ($\eta_{0}$)   & $3\times10^{-4}$ \\
Final learning rate ($\eta_{T}$)     & $1.5\times10^{-4}$ \\
Scheduler                            & Exponential decay \\
Decay factor ($\gamma$)              & $(\eta_{T}/\eta_{0})^{1/50}=0.87$ \\
Weight decay                         & $2\times10^{-2}$ \\
Maximum epochs                       & 50 \\
Early-stopping patience              & 5 epochs \\
\bottomrule
\end{tabular}
\end{table}

\begin{table*}[t]
\centering
\caption{Comparison of prediction accuracy for the encoding model. For each subject, the mean $\pm$ standard error on the test set is reported for the top 5,000 voxels that achieve the highest accuracy on the train data.}
\vspace{1em}
\resizebox{\textwidth}{!}{
\begin{tabular}{lccccccccc}
\toprule
Methods & ridge & gradient & layers & voxel-wise & shared-weights & subj01 & subj02 & subj05 & subj07 \\
\midrule
\multirow{3}{*}{Linear}
 & \checkmark & --          & 1 & \checkmark & -- & 0.501±0.002 & 0.524±0.001 & 0.570±0.001 & 0.421±0.001 \\
 & \checkmark & --          & 1 & --          & \checkmark & 0.500±0.002 & 0.523±0.001 & 0.567±0.001 & 0.420±0.001 \\
 & --          & \checkmark & 1 & --          & \checkmark & 0.484±0.002 & 0.512±0.002 & 0.563±0.001 & 0.405±0.002 \\
\midrule
\multirow{2}{*}{Nonlinear}
 & --          & \checkmark & 2 & --          & \checkmark & 0.492±0.002 & 0.516±0.001 & 0.565±0.001 & 0.411±0.002 \\
 & --          & \checkmark & 3 & --          & \checkmark & 0.491±0.002 & 0.518±0.001 & 0.563±0.001 & 0.411±0.002 \\
\bottomrule
\end{tabular}
}
\label{appendix:encoding_compare}
\end{table*}

\begin{table}[h]
\centering
\caption{Sentence-level prediction accuracy for $\beta$-based versus encoding-model–based models. Values denote mean $\pm$ standard deviation across test images for the top 5{,}000 voxels per subject, selected based on their training-set prediction accuracy.}
\label{appendix:beta_table}
\vspace{0.5em}
\begin{tabular}{lcccc}
\toprule
Condition & subj01 & subj02 & subj05 & subj07 \\
\midrule
w/o Encoding model & 0.201 $\pm$ 0.082 & 0.248 $\pm$ 0.086 & 0.276 $\pm$ 0.096 & 0.094 $\pm$ 0.054 \\
w/ Encoding model  & 0.246 $\pm$ 0.066 & 0.287 $\pm$ 0.075 & 0.306 $\pm$ 0.084 & 0.218 $\pm$ 0.073 \\
\bottomrule
\end{tabular}
\end{table}

\subsubsection{Caption Generation}
\label{appendix:ablation_caption_generation}
We investigate how various hyper-parameter settings influence the accuracy of the voxel captions (Tables~\ref{appendix:beta_table},~\ref{appendix:ablation_table}). Unless otherwise specified, we adopt the primary-analysis configuration: CLIP-Vision as the default contrastive VLM for feature extraction, 50 optimal images per voxel, MiniCPM-V as the MLLM for captioning the optimal images, five extracted keywords, \textit{gpt-4o} as the keyword extraction model, exemplar-based prompting as the prompting strategy, and use of the Sentence Composer.

\paragraph{Effect of the Encoding Model}
\label{appendix:beta_baseline}
Our goal in LaVCa is to characterise voxel selectivity at the level of DNN-derived visual features, rather than to interpret voxels directly from raw $\beta$-values. Nevertheless, to verify that the encoding model indeed improves interpretability—by suppressing noise and non-visual components in the $\beta$-responses and enabling better generalisation to novel stimuli—we performed an additional control analysis using a purely $\beta$-driven baseline.

To directly assess whether the encoding model is necessary for identifying optimal images, we conducted a control analysis in which, for each voxel, we selected the NSD images that elicited the strongest raw $\beta$-values and generated voxel captions from their MS-COCO captions. Importantly, in this baseline the encoding-model weights used in the later stages—keyword filtering and the Sentence Composer—are replaced with a uniform weighting scheme by setting all weights to~1. This ensures that the entire pipeline operates without any feature-level information from the encoding model, providing an interpretation grounded solely in the measured $\beta$-values.

This $\beta$-based approach produced substantially lower predictive accuracy compared to LaVCa (Table~\ref{appendix:beta_table}). Because this baseline relies solely on raw $\beta$-values, it is highly sensitive to measurement noise and non-visual components of brain activity, resulting in captions that generalised poorly to held-out images.

\begin{table*}[t]
\centering
\caption{Comparison of sentence-level brain activity prediction accuracy using different hyperparameters. Accuracy is reported as the mean $\pm$ standard deviation for the top 5,000 voxels in the test data, selected by training-set accuracy.}
\vskip 0.15in
\resizebox{\textwidth}{!}{
\begin{tabular}{lp{2.5cm}cccc}
\toprule
Parameter & Setting & subj01 & subj02 & subj05 & subj07 \\
\midrule
\multirow{4}{*}{Contrastive VLM} 
& SigLIP2     & 0.232$\pm$0.065 & 0.275$\pm$0.075 & 0.294$\pm$0.083 & 0.206$\pm$0.070 \\
& FG-CLIP     & 0.244$\pm$0.070 & 0.285$\pm$0.076 & 0.306$\pm$0.086 & 0.214$\pm$0.074 \\
& CLIP-Text   & 0.246$\pm$0.067 & 0.281$\pm$0.074 & 0.309$\pm$0.084 & 0.216$\pm$0.071 \\
& CLIP-Vision & 0.246$\pm$0.066 & 0.287$\pm$0.075 & 0.306$\pm$0.084 & 0.218$\pm$0.073 \\
\midrule
\multirow{4}{*}{\# optimal images} 
& 5   & 0.239$\pm$0.068 & 0.279$\pm$0.073 & 0.294$\pm$0.083 & 0.209$\pm$0.073 \\
& 10  & 0.243$\pm$0.068 & 0.281$\pm$0.072 & 0.297$\pm$0.083 & 0.212$\pm$0.074 \\
& 50  & 0.246$\pm$0.066 & 0.287$\pm$0.075 & 0.306$\pm$0.084 & 0.218$\pm$0.073 \\
& 100 & 0.246$\pm$0.067 & 0.285$\pm$0.074 & 0.301$\pm$0.083 & 0.215$\pm$0.072 \\
\midrule
\multirow{2}{*}{Multimodal LLM}
  & MiniCPM-V & 0.246$\pm$0.066 & 0.287$\pm$0.075 & 0.306$\pm$0.084 & 0.218$\pm$0.073 \\
  & BLIP      & 0.242$\pm$0.068 & 0.285$\pm$0.075 & 0.302$\pm$0.084 & 0.215$\pm$0.072 \\ 
\midrule
\multirow{3}{*}{\# keywords} 
& 1  & 0.237$\pm$0.066 & 0.274$\pm$0.073 & 0.295$\pm$0.085 & 0.207$\pm$0.072 \\
& 5  & 0.246$\pm$0.066 & 0.287$\pm$0.075 & 0.306$\pm$0.084 & 0.218$\pm$0.073 \\
& 10 & 0.241$\pm$0.067 & 0.279$\pm$0.074 & 0.296$\pm$0.084 & 0.212$\pm$0.074 \\
\midrule
\multirow{3}{*}{Extraction model} 
& TextGraphParser & 0.242$\pm$0.067 & 0.276$\pm$0.073 & 0.296$\pm$0.084 & 0.205$\pm$0.071 \\
& Llama3.1-70B    & 0.238$\pm$0.067 & 0.281$\pm$0.075 & 0.298$\pm$0.085 & 0.214$\pm$0.073 \\
& gpt-4o          & 0.246$\pm$0.066 & 0.287$\pm$0.075 & 0.306$\pm$0.084 & 0.218$\pm$0.073 \\
\midrule
\multirow{2}{*}{Prompt}
  & Hidden CoT & 0.239$\pm$0.064 & 0.283$\pm$0.074 & 0.298$\pm$0.083 & 0.214$\pm$0.072 \\
  & Exemplar-based & 0.246$\pm$0.066 & 0.287$\pm$0.075 & 0.306$\pm$0.084 & 0.218$\pm$0.073 \\
\midrule
\multirow{2}{*}{Sentence Composer}
  & \ding{51}  & 0.246$\pm$0.066 & 0.287$\pm$0.075 & 0.306$\pm$0.084 & 0.218$\pm$0.073 \\
  & \ding{55}  & 0.230$\pm$0.066 & 0.279$\pm$0.078 & 0.296$\pm$0.087 & 0.201$\pm$0.070 \\
\bottomrule
\end{tabular}
}
\label{appendix:ablation_table}
\end{table*}

\paragraph{Contrastive VLM comparison.}
We next compare different contrastive vision–language models (VLMs) used to extract latent features for voxel-wise encoding. 
Specifically, we evaluate SigLIP2~\citep{tschannen2025siglip}, FG-CLIP~\citep{xie2025fg}, and both the text and vision branches of CLIP~\citep{radford2021learning}. 
For CLIP-Text, we obtain projection-layer embeddings from the COCO captions that were pre-assigned to the NSD image stimuli. 
Overall, CLIP-based representations (CLIP-Text and CLIP-Vision) achieve the highest accuracies.
FG-CLIP performs comparably to CLIP-Vision, while SigLIP2 does not surpass CLIP in our setting—despite reports in the original SigLIP work that it often outperforms CLIP on benchmark tasks—yet its inclusion demonstrates that LaVCa generalises well across diverse VLM backbones.

\paragraph{Number of optimal images.}
We vary the number of optimal images used for keyword extraction from 5 to 10, 50, and 100.  
Increasing the number up to 50 improves accuracy, presumably because relying only on top-ranked images can omit useful second- and third-ranked keywords. Using more images therefore captures a broader range of selective concepts.  
However, once the number of optimal images reaches 100, the improvement plateaus, likely because additional concepts can no longer be adequately represented with only five keywords.  
These observations suggest that increasing the number of keywords, rather than merely adding more images, may further enhance accuracy.

\paragraph{Multimodal LLM comparison.}
We compare two multimodal LLMs for captioning the optimal images: the state-of-the-art MiniCPM-V and the lighter, less accurate BLIP.  
MiniCPM-V slightly outperforms BLIP, indicating that a more capable MLLM can further improve LaVCa’s voxel-caption accuracy.  
Conversely, the modest gap between BLIP and MiniCPM-V suggests that our approach generalises well even with simpler captioning models.

\paragraph{Number of extracted keywords.}
With the number of optimal images fixed at 50, we vary the number of extracted keywords among 1, 5, and 10.  
Increasing the output concepts from one to five boosts accuracy, whereas extending the list to ten decreases accuracy—likely because irrelevant or noisy concepts are introduced.  
The improvement at five keywords indicates that voxels encode multiple concepts, but extracting too many can introduce noise.  
Thus, expanding the image set rather than the keyword count may be a more effective strategy for capturing additional informative concepts.

\paragraph{Keyword-extraction model comparison.}
We evaluate three models for extracting keywords from the optimal image set: \textit{gpt-4o}, an 8-bit-quantised Llama 3.1-70B-Instruct, and the TextGraphParser \cite{li2023factual} employed in MeaCap.  
\textit{gpt-4o} surpasses TextGraphParser, showing that an open-ended LLM makes concept extraction more effective than simply pulling words from captions.  
It also exceeds Llama 3.1-70B-8bit, demonstrating that stronger LLMs can further raise accuracy.  
These results imply that LaVCa’s interpretability will improve in tandem with future advances in LLM capability.

\paragraph{Prompt comparison.}
We compare two prompting strategies for summarising the captions associated with each voxel’s optimal image set: \textit{Exemplar-based Prompting} and a more structured \textit{Hidden CoT Prompting} formulation.  
Exemplar-based Prompting follows the classical in-context learning paradigm~\citep{brown2020language,dunlap2024describing}, where multiple positive and negative examples constrain the output format and guide the model toward producing concept-like captions.  
In contrast, the structured prompt incorporates an expert role assignment and an internal (“hidden”) chain-of-thought instruction~\citep{wei2022chain}, encouraging the model to silently derive recurrent patterns before producing the final set of concepts.

Across subjects, both prompting strategies performed similarly, indicating that LaVCa is robust to prompt specification.  
Nonetheless, Exemplar-based Prompting consistently achieved slightly higher accuracies for all subjects, suggesting that explicit examples remain an effective mechanism for stabilising the output structure in this caption-concept abstraction task.

\paragraph{Effect of the Sentence Composer.}
Finally, we assess the Sentence Composer by comparing results with and without it.  
Incorporating the Sentence Composer yields higher accuracy than relying on keywords alone, suggesting that contextual information beyond isolated concepts enables a more fine-grained interpretation of voxel properties.

\begin{table*}[t]
\centering
\caption{Comparison of sentence similarity models for accuracy evaluation. Accuracy is reported as the mean $\pm$ standard deviation for the top 5,000 voxels in the test data, selected by training-set accuracy.}
\vskip 0.15in
\begin{tabular}{lcccc}
\toprule
Model & subj01 & subj02 & subj05 & subj07 \\
\midrule
MPNet  & 0.245$\pm$0.067 & 0.287$\pm$0.075 & 0.305$\pm$0.088 & 0.216$\pm$0.069 \\
Sentence-BERT          & 0.246$\pm$0.066 & 0.287$\pm$0.075 & 0.306$\pm$0.084 & 0.218$\pm$0.073 \\
\bottomrule
\end{tabular}
\label{appendix:sentence_similarity}
\end{table*}

\subsubsection{Sentence similarity model for evaluation}
\label{appendix:ablation_sentence_similarity}
We also examine how the choice of sentence similarity model used for evaluation affects voxel-caption accuracy (Table~\ref{appendix:sentence_similarity}). 
Specifically, we compare MPNet~\citep{song2020mpnet} and Sentence-BERT~\citep{reimers2019sentence}, both widely used models for computing sentence embeddings. 
Overall, the two models yield nearly identical accuracies across all subjects, with Sentence-BERT performing marginally better. 
This consistency suggests that LaVCa’s evaluation results are robust to the particular choice of sentence similarity model, and that the observed improvements are not dependent on model-specific idiosyncrasies.

\begin{figure*}[t] 
  \centering
  \includegraphics[width=\textwidth]{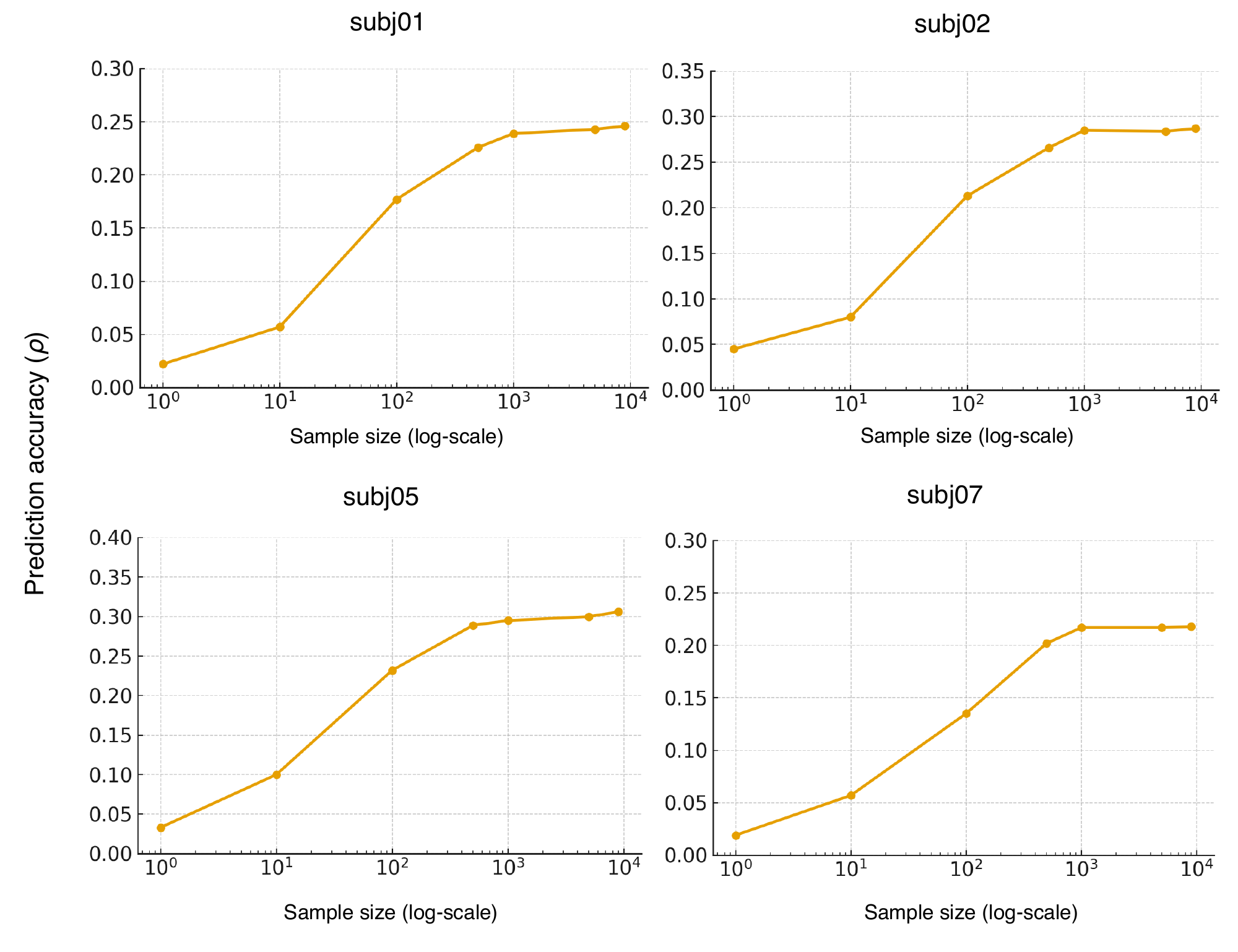}
  \caption{Effect of sample size on test prediction accuracy. Mean prediction accuracy on the test data for the top 5000 voxels selected using training-data performance, plotted across different sample sizes (x-axis in log scale).
}
  \label{appendix:scaling_sample_size}
\end{figure*}

\subsection{Data-size Sensitivity}
\label{appendix:data-size-sensitivity}
In this appendix, we examine how LaVCa’s caption-prediction accuracy depends on the amount and structure of the stimulus data used within the NSD dataset. We vary (i) the number of stimulus images supplied to LaVCa and (ii) the number of stimulus-image categories used as part of the method’s categorization stage.

\subsubsection{Sample-size Sensitivity}
To evaluate how the size of the NSD dataset influences LaVCa’s performance, we vary the number of stimulus images provided to the method across 1, 10, 100, 500, 1000, 5000, and approximately 9000 (the full NSD stimulus set).  
Figure~\ref{appendix:scaling_sample_size} summarizes how prediction accuracy changes across these sample sizes.

The number of stimulus-image categories is fixed at 80, based on the COCO object categories, except when fewer than 80 images are used, in which case the number of categories matches the sample size.

LaVCa’s accuracy shows a clear dependence on the number of available stimulus images. With only a small number of images (such as 1 or 10), accuracy decreases substantially because the limited variety of stimuli restricts the method’s ability to infer stable relationships between fMRI responses and visual content. As more images are incorporated, accuracy improves rapidly and stabilizes once approximately 500 to 1000 stimulus images are available. Increasing the stimulus dataset beyond this range yields only modest gains, and the performance with the full \textasciitilde9000-image set is only slightly higher than that achieved with 5000 images. These results indicate that roughly 500–1000 stimulus images are sufficient for LaVCa to produce reliable voxel-level captions, without requiring the full scale of the NSD stimulus set.

\begin{figure*}[t] 
  \centering
  \includegraphics[width=\textwidth]{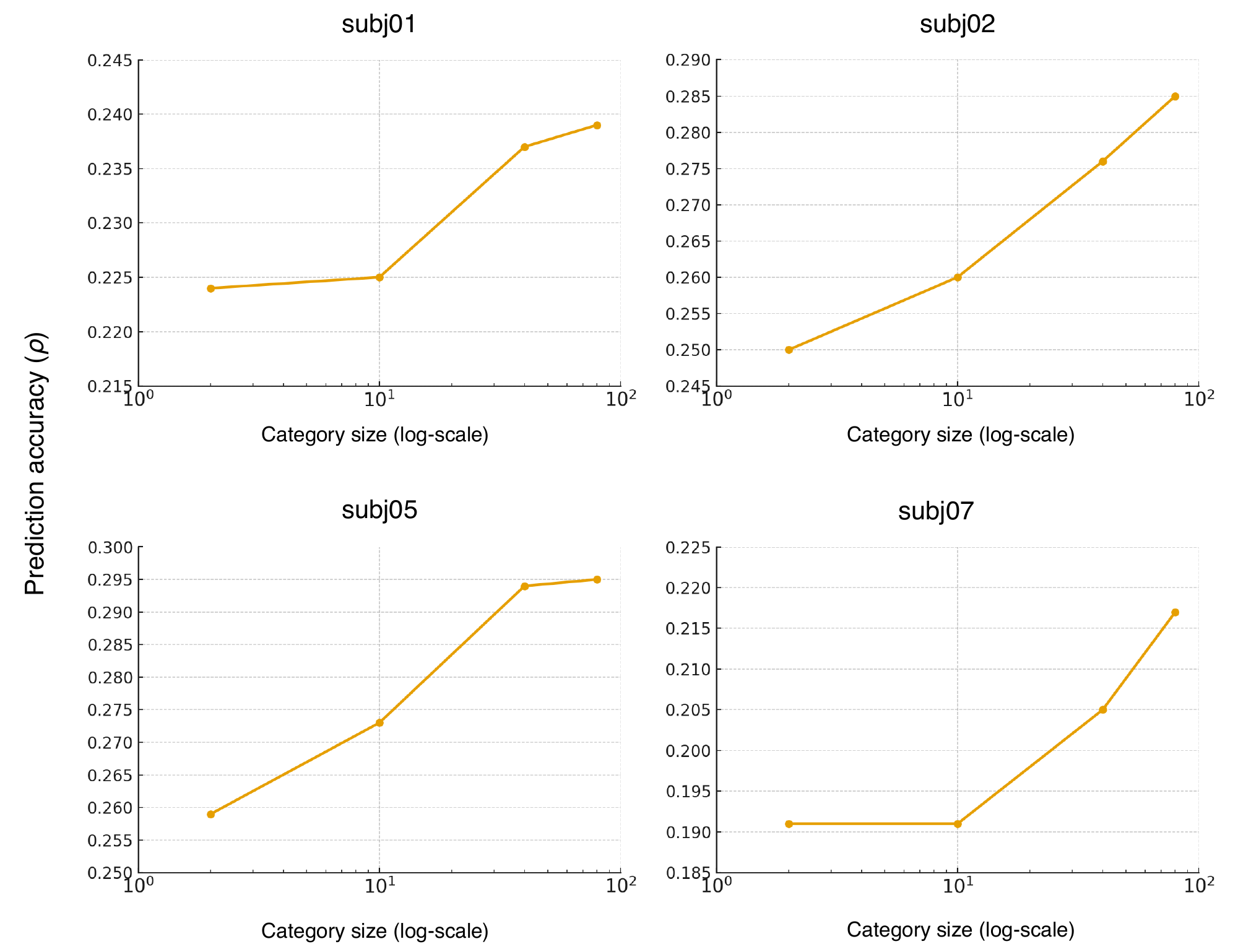}
  \caption{Effect of category size on test prediction accuracy. Mean prediction accuracy on the test data for the top 5000 voxels selected using training-data performance, shown for different category sizes.
}
  \label{appendix:scaling_category_size}
\end{figure*}

\subsubsection{Category-size Sensitivity}
We next analyze how the number of stimulus-image categories provided to LaVCa affects caption-prediction accuracy.  
Figure~\ref{appendix:scaling_category_size} provides a summary of this analysis.  
The number of categories is varied across 2, 10, 40, and 80 only during the training stage, while the test evaluation uses the full set of stimulus images.

LaVCa’s accuracy increases as the categories become more fine-grained. Extremely coarse category sets, such as those with 2 or 10 categories, compress variability across the stimulus set and lead to reduced accuracy. Using 40 categories captures substantially more structure within the stimulus data and already approaches the performance of the full 80-category condition. The 80-category condition performs best, particularly in higher-level visual areas where finer distinctions between visual stimulus types are beneficial.

\subsection{Additional fMRI Dataset: BOLD5000}
\label{appendix:additional_fmri_dataset}
\subsubsection{Dataset and preprocessing}

To assess whether LaVCa generalises beyond NSD, we additionally evaluate our method on the BOLD5000 dataset, which contains fMRI responses from four subjects viewing approximately 5{,}000 natural images~\citep{chang2019bold5000}.
Following our NSD pipeline, we first compute single-trial response patterns using GLMsingle $\beta$-estimates for each session and concatenate them along the time dimension to obtain a 4D volume for each subject.  

For our analyses, we restrict all experiments to the functional masks provided in BOLD5000, consisting of EarlyVis (early visual cortex), LOC (lateral occipital complex), OPA (occipital place area), PPA (parahippocampal place area), and RSC (retrosplenial cortex).  
Across these regions, the total number of voxels amounts to approximately 2{,}500 per subject.
Within each subject, we vectorise the GLMsingle $\beta$-estimates into trial-wise response matrices and standardise each voxel by z-scoring across all trials. Following the same procedure as in NSD, we first split the trials into 90\% for training and validation and 10\% for testing. 

We then apply LaVCa to BOLD5000 using exactly the same configuration as in the main NSD experiments, without any dataset-specific tuning.

Because LaVCa evaluates voxel captions by comparing them with the ground-truth captions of the held-out stimuli, textual captions for the test images are required.  
In BOLD5000, the only subset of stimuli accompanied by human-written captions is the set of COCO images; therefore, test-set accuracy is computed exclusively on COCO trials using their official COCO captions and the corresponding $\beta$ values.  
Importantly, the encoding model is trained on all available BOLD5000 trials, and only the evaluation stage is restricted to COCO trials.

\subsubsection{Results on BOLD5000}

We apply LaVCa to the BOLD5000 dataset and compute voxel-caption accuracy on the COCO-captioned trials in the test split. 
Table~\ref{tab:bold5000_caption_accuracy} summarises the caption-prediction accuracy for the top 100, top 500, and top 1000 voxels in each subject, selected based on their training-set performance.

Across all voxels, LaVCa achieved significant caption-prediction accuracy in roughly 40\% of voxels.  

Although LaVCa’s performance on BOLD5000 is lower than on NSD, this reduction closely parallels the decrease in encoding-model accuracy between the two datasets.  
Specifically, the average encoding-model performance drops from 0.504 in NSD to 0.302 in BOLD5000—an approximately 40\% reduction.  
The proportional decline in both encoding and caption-prediction accuracy suggests that the diminished performance reflects inherent differences in dataset quality (e.g., shorter scan durations and lower SNR in BOLD5000), rather than overfitting of LaVCa to NSD.

Taken together, these results demonstrate that LaVCa generalises to a dataset with different subjects and stimulus sets.

\begin{table*}[t]
\centering
\caption{
Comparison of sentence\mbox{-}level caption prediction accuracy on BOLD5000 (subj01--subj04). 
``Top-\emph{N} voxels'' refers to the voxels with top-\emph{N} prediction performance in the training data.  
Values are mean $\pm$ standard deviation on the test data.
}
\vskip 0.1in
\resizebox{\textwidth}{!}{%
\begin{tabular}{lcccc}
\toprule
\textbf{Top-$N$ voxels} & \textbf{subj01} & \textbf{subj02} & \textbf{subj03} & \textbf{subj04} \\
\midrule
Top-100  & 
$0.2655 \pm 0.0609$ & 
$0.1927 \pm 0.0656$ & 
$0.1525 \pm 0.0540$ & 
$0.2117 \pm 0.0576$ \\
Top-500  & 
$0.1652 \pm 0.0780$ & 
$0.1138 \pm 0.0737$ & 
$0.1185 \pm 0.0560$ & 
$0.1523 \pm 0.0668$ \\
Top-1000 & 
$0.1286 \pm 0.0769$ & 
$0.0776 \pm 0.0749$ & 
$0.1006 \pm 0.0575$ & 
$0.1221 \pm 0.0743$ \\
\bottomrule
\end{tabular}}
\label{tab:bold5000_caption_accuracy}
\end{table*}

\begin{figure*}[t] 
  \centering
  \includegraphics[width=\textwidth]{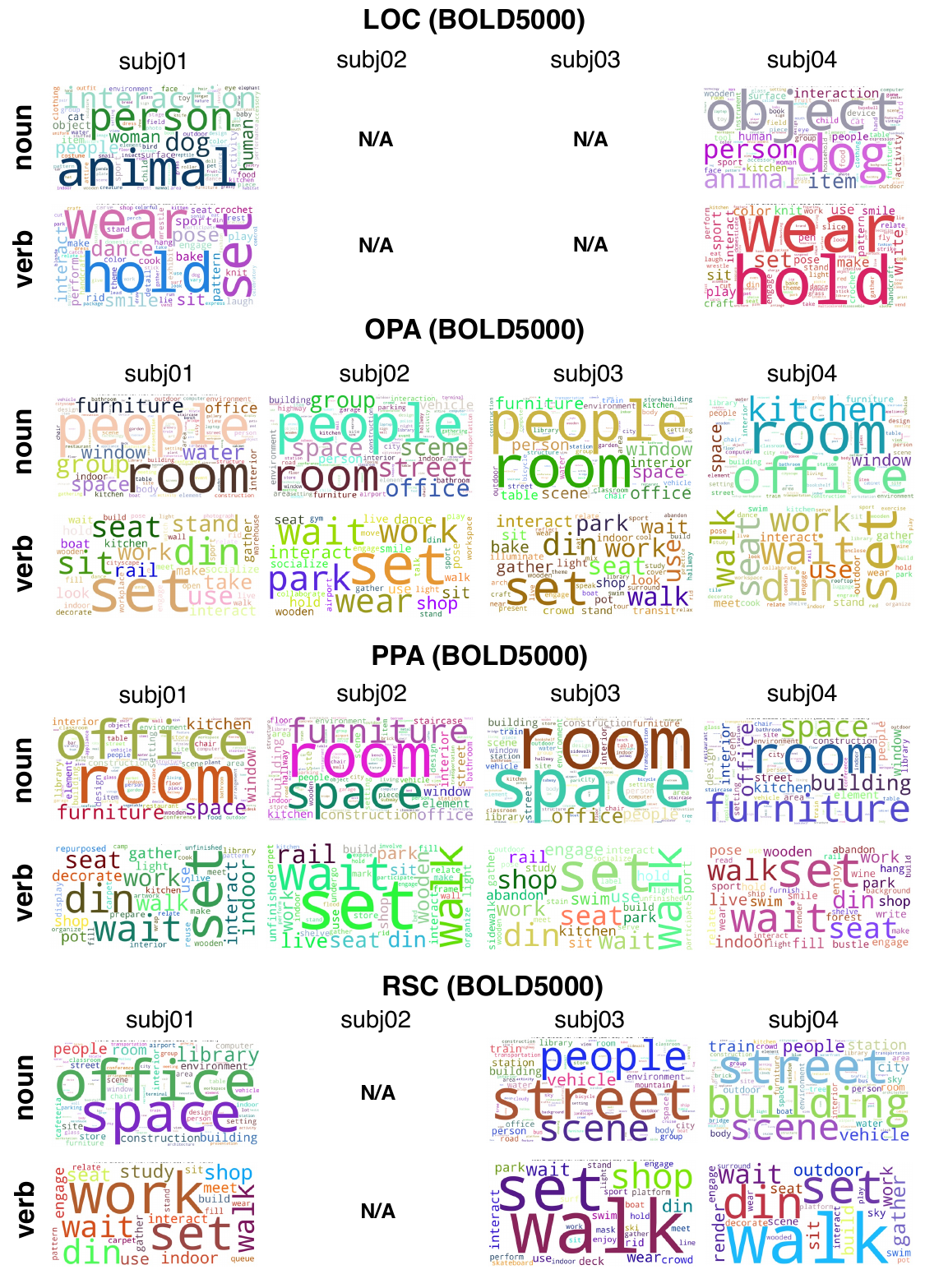}
  \caption{
Word clouds depicting the 100 most frequent nouns and verbs appearing in voxel captions for the BOLD5000 dataset. Each panel shows the word cloud for a particular brain region and subject.
}
  \label{appendix:bold5000_wordcloud_all}
\end{figure*}

To assess the reproducibility of voxel-level semantic organization across datasets and subject groups, we conducted a WordCloud-based analysis analogous to that used in the NSD experiments (Figure~\ref{appendix:bold5000_wordcloud_all}). 

Across ROIs, the BOLD5000 results exhibit semantic patterns that closely parallel those observed in NSD despite differences in subjects, scanner parameters, and stimulus sets. In BOLD5000 LOC, which is broadly associated with object- and body-related processing, we observed frequent person-related nouns such as \textit{person}, reproducing the person-related cluster identified in NSD's EBA. Verb distributions show a similar correspondence: BOLD5000 LOC contained action verbs such as \textit{wear} and \textit{hold}, mirroring the body- and interaction-related action clusters prominent in NSD's EBA and OFA.

Scene-selective regions demonstrated an even stronger cross-dataset alignment. BOLD5000 OPA, PPA, and RSC all exhibited WordClouds enriched with place-related nouns including \textit{room}, \textit{kitchen}. These terms match the dominant scene-related cluster observed in NSD PPA, indicating that the underlying semantic structure of these ROIs is stable across datasets. 

Taken together, these results show that the semantic clusters identified by LaVCa are not specific to a particular dataset. Instead, similar diversity patterns emerge across both NSD and BOLD5000: object- and body-related concepts appear in lateral occipito-temporal regions, and place-related concepts are consistently represented in medial scene-selective regions. This cross-dataset reproducibility directly addresses the reviewer’s question regarding the stability of semantic clusters across large populations and distinct image sets.

\subsection{Limitaion}
\label{appendix:limitation}
Despite the overall improvement in brain activity prediction, we observe that face-selective regions do not achieve accuracy levels as high as those in other ROIs (Figure \ref{results:ROI_caption_shuffle}). One reason may be that our current approach uses a Multimodal LLM (MLLM) to produce relatively simple captions for optimal images, often omitting important local features (e.g., ``eyes,'' ``nose'') and focusing on more global terms (e.g., ``face,'' ``person''). Consequently, the subsequent summarization step lacks access to these local details. Because our method relies on language descriptions, it has inherent limitations in capturing the fine-grained, local selectivity of these voxels. Incorporating recent techniques that visually interpret local voxel selectivity ~\citep{luo2024brain} could help address this gap.

Furthermore, while our current study describes voxel selectivity primarily in response to visual stimuli in the occipital cortex, there exist ``multimodal voxels'' in the brain that are simultaneously activated by auditory and linguistic information, and higher-order cognitive processes such as calculations, memory retrieval, and reasoning ~\citep{nakai2020quantitative, nakai2022representations}. Designing stimuli and experimental tasks encompassing diverse sensory inputs (e.g., auditory, textual) and cognitive challenges (e.g., recalling past events, performing reasoning tasks) is essential when interpreting such voxels. Because our approach uses LLM-based textual summarization, it can be adapted to represent a wide range of stimuli and cognitive states in text form, providing a unified framework for multimodal integration. Looking ahead, by jointly modeling images, semantic information, auditory features, and cognitive tasks, we anticipate capturing the brain’s integrated representation of both sensory and higher-order cognitive functions with greater accuracy.

\subsection{Impact Statement}
\label{appendix:impact_statement}
We introduce a data-driven method that uses a large language model to generate natural language captions of voxel-level visual selectivity. Using the method detailed in this paper, we aim to provide a more fine-grained understanding of human visual function than previously achieved. We acknowledge that this human brain research could raise concerns regarding individual privacy. Although the present study examined relatively coarse-grained voxel-level data, we cannot dismiss the possibility that future advances in measurement and analysis techniques may enable the extraction of more detailed individual-specific information. In any case, obtaining explicit informed consent from participants remains crucial when collecting and using human brain activity data, as with the NSD dataset used in this study.

\clearpage

\begin{figure*}[t] 
  \centering
  \includegraphics[width=\textwidth]{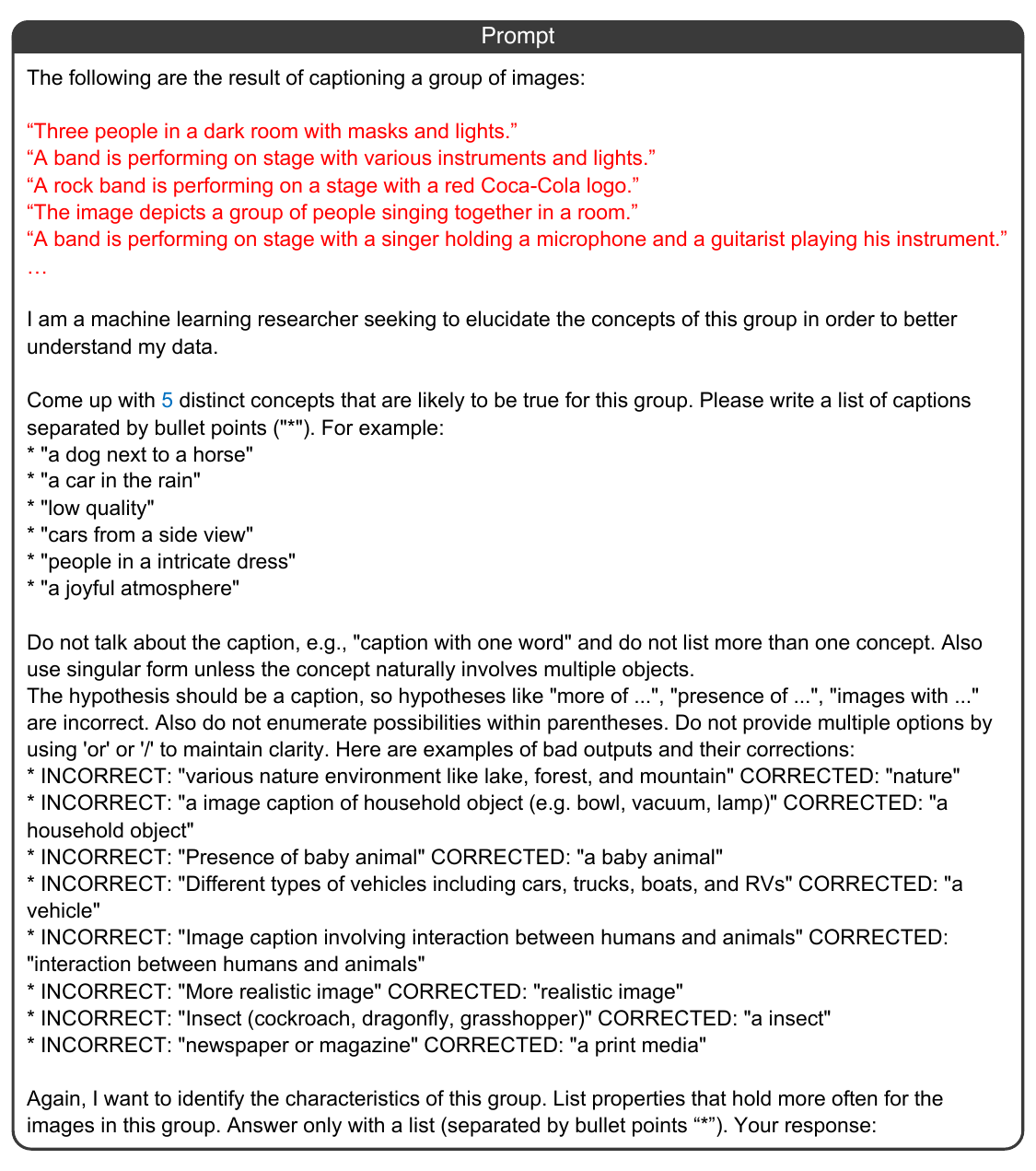}
\caption{
The prompt used for summarizing the captions of optimal image groups with an LLM (Exemplar-based Prompting). 
The text in red represents the captions of the optimal image group, which 
depend on the target voxel and the number of optimal images used.  
The blue number specifies the required number of concepts, which was varied during the ablation study.
}
  \label{appendix:prompt}
\end{figure*}

\begin{figure*}[t] 
  \centering
  \includegraphics[width=\textwidth]{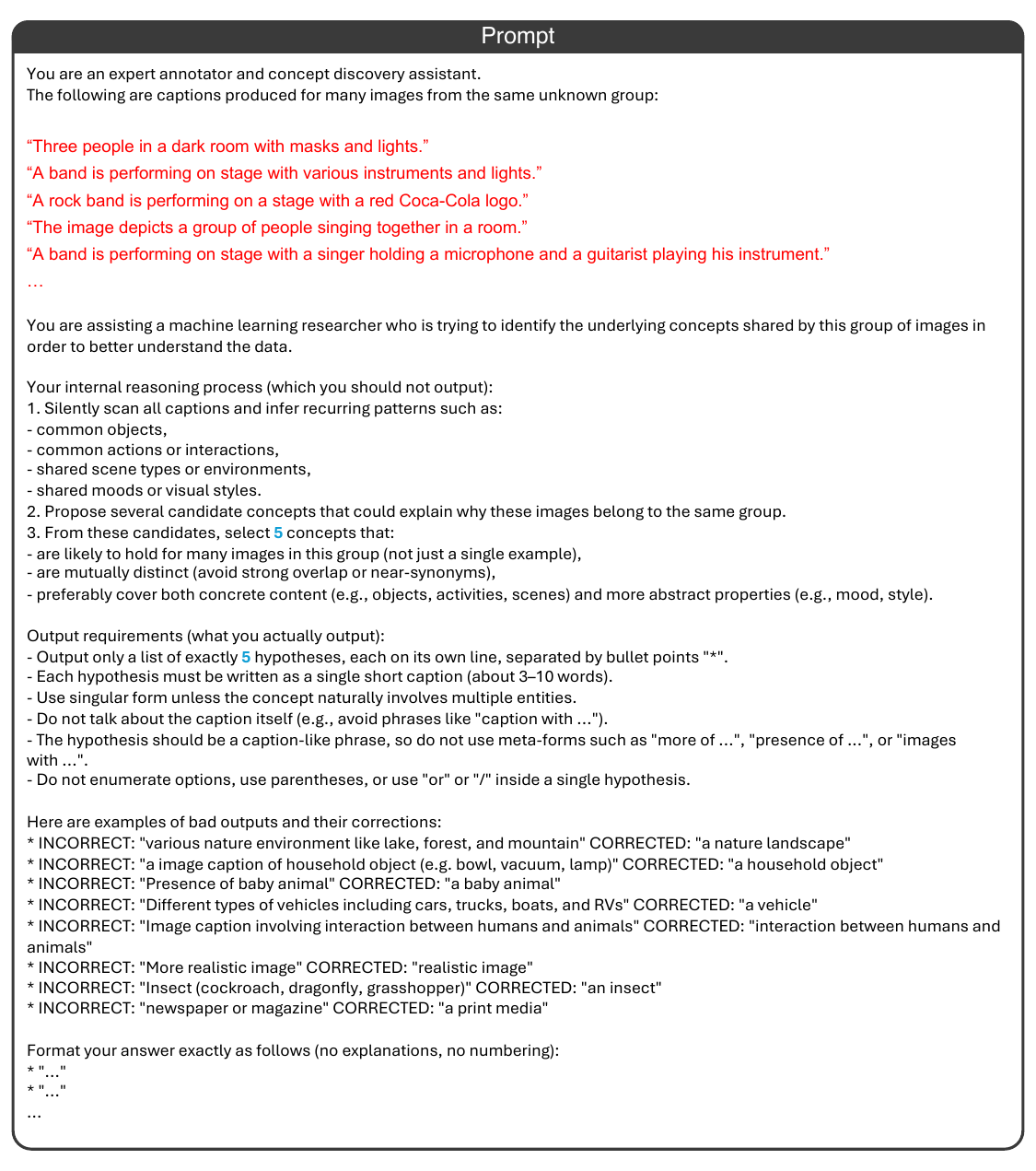}
\caption{
The prompt used for summarizing the captions of optimal image groups with an LLM (Hidden CoT Prompting).
The text in red represents the captions of the optimal image group, which 
depend on the target voxel and the number of optimal images used.  
The blue number specifies the required number of concepts, which was varied during the ablation study.
}
  \label{appendix:prompt}
\end{figure*}

\begin{figure*}[t]
  \centering
  \includegraphics[width=\textwidth]{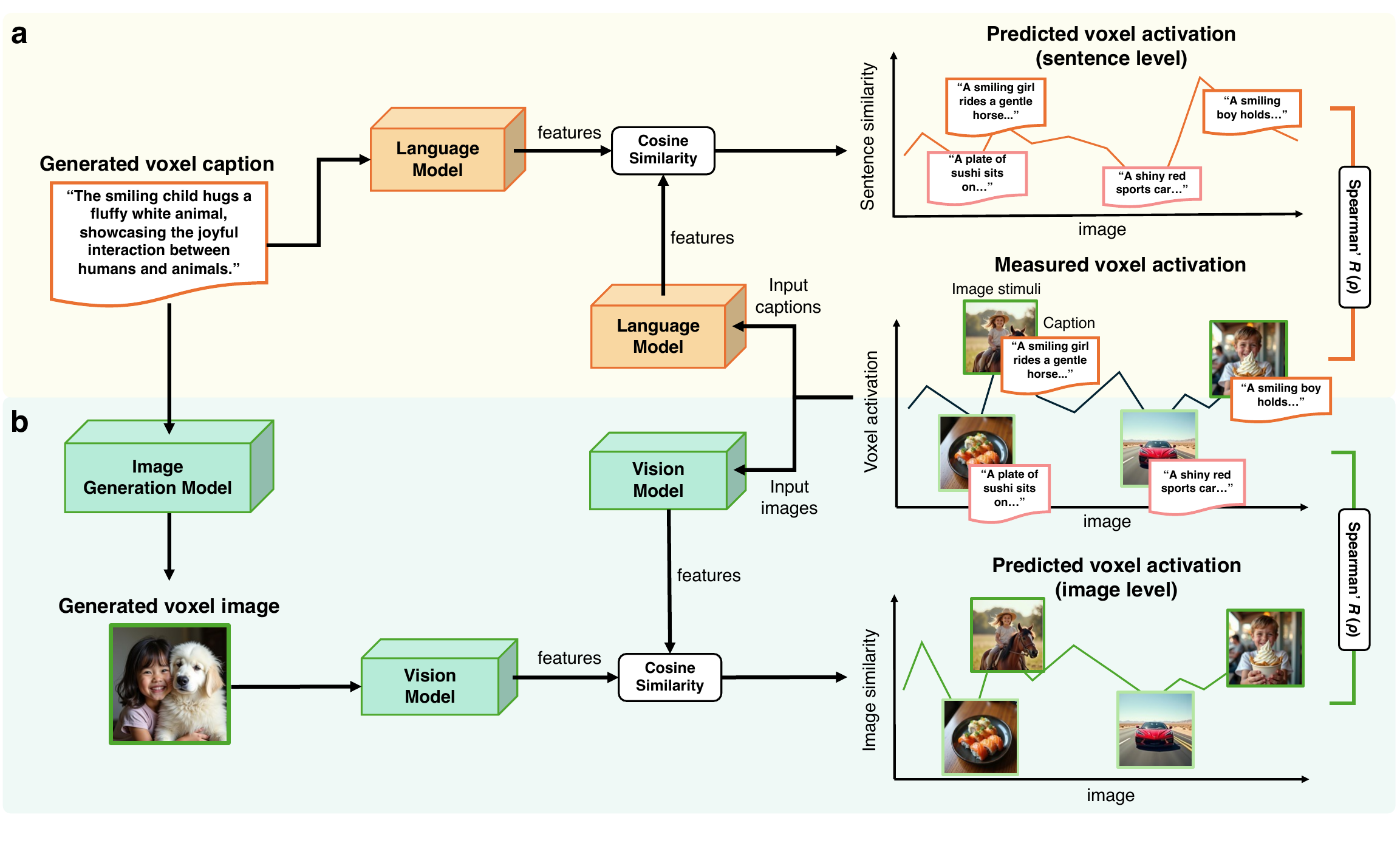}
  \caption{Overview of our evaluation methods. (a) We first obtain text embeddings for both the generated voxel captions (from the language model) and the NSD image captions. We then compute the cosine similarity between the voxel caption embeddings and each NSD caption embedding to derive a rough prediction of voxel activity. Finally, we evaluate text-level prediction performance by calculating Spearman’s rank correlation coefficient between these predicted values and the actual voxel responses. (b) We generate voxel images by visualizing voxel captions with an image generation model and, using the same vision model, compute vision-based embeddings for both the generated voxel images and the NSD image stimuli. As in (a), we compute the cosine similarity between voxel-image embeddings and NSD image embeddings and use Spearman’s rank correlation coefficient to evaluate image-level prediction performance.}
  \label{appendix:voxel_pred_method}
\end{figure*}

\begin{figure*}[t] 
  \centering
  \includegraphics[width=0.8\textwidth]{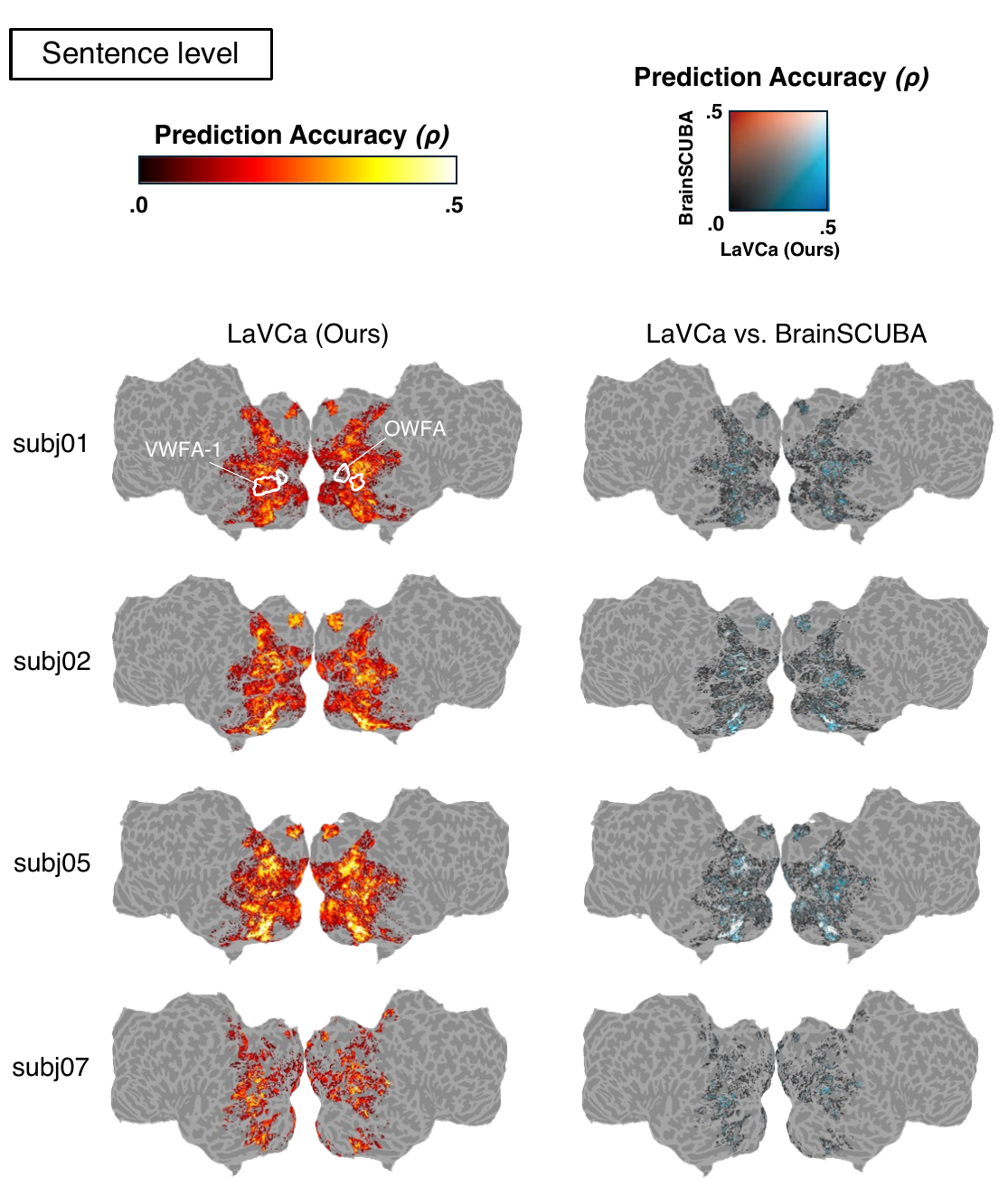}
  \caption{Mapping of brain activity prediction accuracy at the sentence level for LaVCa (left) and a comparison of brain activity prediction accuracy at the sentence level between LaVCa and BrainSCUBA (right) onto the flatmap for all subjects. The white outlines indicate Visual Word Form Area (VWFA-1) and Occipital Word Form Area (OWFA), which are ranked among the top two Words-category ROIs based on the mean number of voxels across subjects.
}
  \label{appendix:sentence_cc_flatmap}
\end{figure*}

\begin{figure*}[t] 
  \centering
  \includegraphics[width=0.8\textwidth]{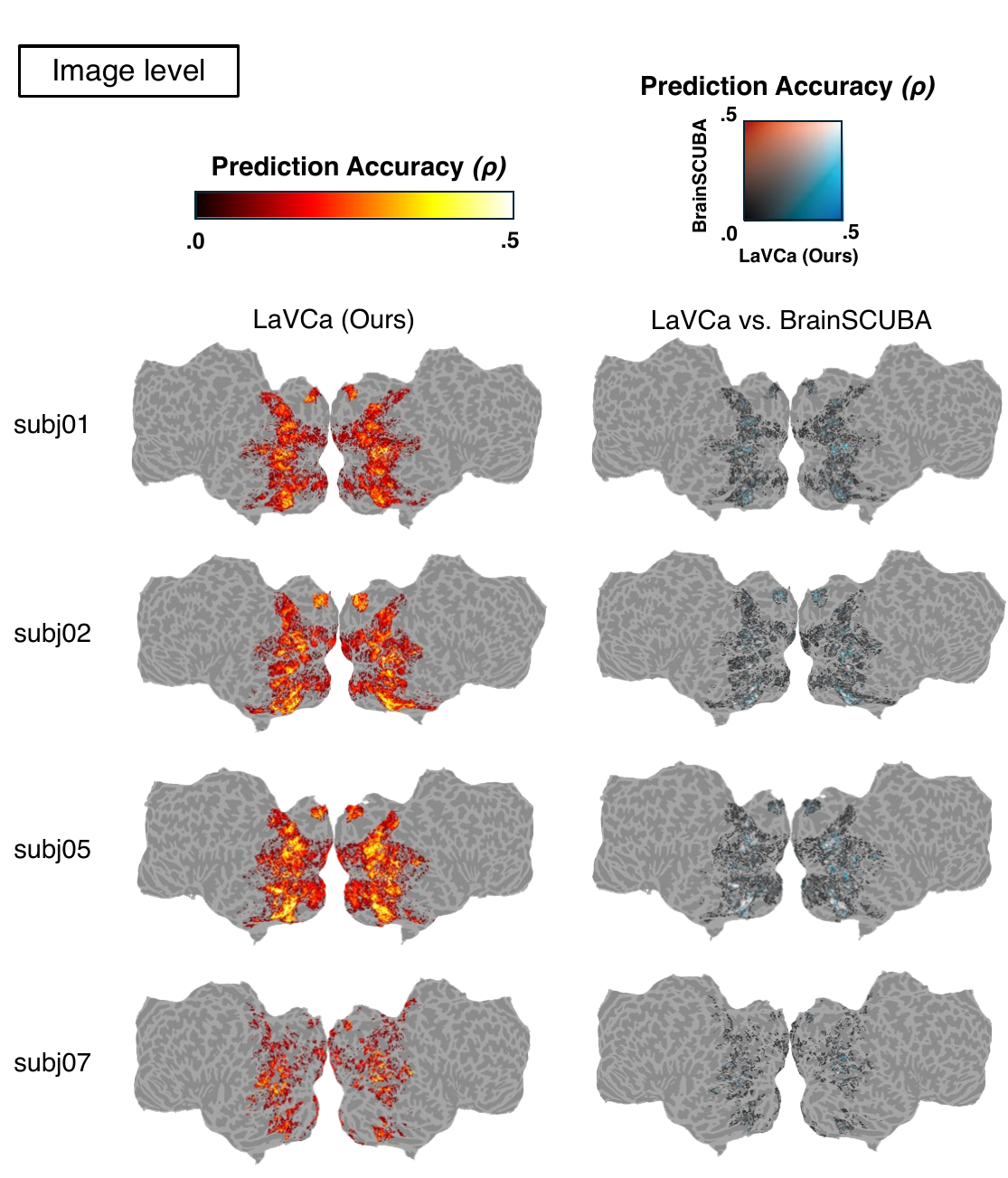}
  \caption{Mapping of brain activity prediction accuracy at the image level for LaVCa (left) and a comparison of brain activity prediction accuracy at the image level between LaVCa and BrainSCUBA (right) onto the flatmap for all subjects.
}
  \label{appendix:image_cc_flatmap}
\end{figure*}

\clearpage
\begin{table*}[t]
\centering
\caption{Comparison of sentence-level brain activity prediction performance (all subjects). “Top-$N$ voxels’’ refers to the voxels with top-$N$ prediction performance in the training data.  Each cell shows the mean ± standard deviation of prediction performance on the test data.  Two additional columns indicate the hyper-parameter setting of LaVCa variants.}
\vskip 0.1in
\definecolor{lavcabg}{HTML}{B5EAD7} 
\resizebox{\textwidth}{!}{%
\begin{tabular}{lcccccc}
\toprule
\multicolumn{7}{c}{\textbf{Top1000 voxels}}\\
\midrule
Model & \# keywords & Sentence Composer & subj01 & subj02 & subj05 & subj07\\
\midrule
Shuffled & -- & -- & -0.010 ± 0.281 & 0.094 ± 0.312 & 0.145 ± 0.331 & 0.013 ± 0.265 \\
BrainSCUBA & -- & -- & 0.291 ± 0.049 & 0.347 ± 0.062 & 0.378 ± 0.068 & 0.267 ± 0.055 \\
LaVCa (Ours) & 1 & \ding{55} & 0.300 ± 0.054 & 0.352 ± 0.057 & 0.393 ± 0.059 & 0.286 ± 0.056 \\
\rowcolor{lavcabg}\textbf{LaVCa (Ours)} & 5 & \ding{51} & \textbf{0.338 ± 0.051} & \textbf{0.392 ± 0.057} & \textbf{0.420 ± 0.061} & \textbf{0.320 ± 0.060} \\
\midrule
\multicolumn{7}{c}{\textbf{Top3000 voxels}}\\
\midrule
Model & \# keywords & Sentence Composer & subj01 & subj02 & subj05 & subj07\\
\midrule
Shuffled & -- & -- & 0.000 ± 0.228 & 0.059 ± 0.255 & 0.099 ± 0.274 & 0.004 ± 0.205 \\
BrainSCUBA & -- & -- & 0.237 ± 0.057 & 0.284 ± 0.067 & 0.305 ± 0.077 & 0.212 ± 0.061 \\
LaVCa (Ours) & 1 & \ding{55} & 0.240 ± 0.062 & 0.288 ± 0.068 & 0.317 ± 0.077 & 0.221 ± 0.067 \\
\rowcolor{lavcabg}\textbf{LaVCa (Ours)} & 5 & \ding{51} & \textbf{0.280 ± 0.059} & \textbf{0.325 ± 0.068} & \textbf{0.349 ± 0.075} & \textbf{0.253 ± 0.069} \\
\midrule
\multicolumn{7}{c}{\textbf{Top5000 voxels}}\\
\midrule
Model & \# keywords & Sentence Composer & subj01 & subj02 & subj05 & subj07\\
\midrule
Shuffled & -- & -- & 0.007 ± 0.199 & 0.058 ± 0.223 & 0.067 ± 0.243 & 0.009 ± 0.175 \\
BrainSCUBA & -- & -- & 0.207 ± 0.062 & 0.251 ± 0.071 & 0.264 ± 0.084 & 0.182 ± 0.065 \\
LaVCa (Ours) & 1 & \ding{55} & 0.205 ± 0.068 & 0.250 ± 0.075 & 0.272 ± 0.086 & 0.186 ± 0.072 \\
\rowcolor{lavcabg}\textbf{LaVCa (Ours)} & 5 & \ding{51} & \textbf{0.246 ± 0.066} & \textbf{0.287 ± 0.075} & \textbf{0.306 ± 0.084} & \textbf{0.218 ± 0.073} \\
\midrule
\multicolumn{7}{c}{\textbf{Top10000 voxels}}\\
\midrule
Model & \# keywords & Sentence Composer & subj01 & subj02 & subj05 & subj07\\
\midrule
Shuffled & -- & -- & 0.008 ± 0.157 & 0.039 ± 0.178 & 0.051 ± 0.192 & 0.012 ± 0.134 \\
BrainSCUBA & -- & -- & 0.159 ± 0.071 & 0.195 ± 0.081 & 0.199 ± 0.095 & 0.134 ± 0.072 \\
LaVCa (Ours) & 1 & \ding{55} & 0.154 ± 0.076 & 0.190 ± 0.086 & 0.199 ± 0.101 & 0.132 ± 0.080 \\
\rowcolor{lavcabg}\textbf{LaVCa (Ours)} & 5 & \ding{51} & \textbf{0.191 ± 0.077} & \textbf{0.227 ± 0.086} & \textbf{0.237 ± 0.098} & \textbf{0.163 ± 0.081} \\
\bottomrule
\end{tabular}}
\label{appendix:TopN_sentence_acc_comparison}
\end{table*}

\clearpage
\begin{table*}[t]
\centering
\caption{Comparison of image\mbox{-}level brain activity prediction performance (all subjects). 
"Top-\emph{N} voxels" refers to the voxels with top-\emph{N} prediction performance in the training data.
Values are mean ± standard deviation on the test data.}
\vskip 0.1in
\definecolor{lavcabg}{HTML}{B5EAD7}

\resizebox{\textwidth}{!}{%
\begin{tabular}{lllllll}
\toprule
\multicolumn{7}{c}{\textbf{Top1000 voxels}} \\
\midrule
Model & \# keywords & Sentence Composer & subj01 & subj02 & subj05 & subj07 \\
\midrule
Shuffled & -- & -- & 0.022 ± 0.235 & 0.048 ± 0.254 & 0.104 ± 0.273 & 0.036 ± 0.230 \\
BrainSCUBA & -- & -- & 0.278 ± 0.056 & 0.322 ± 0.052 & 0.357 ± 0.057 & 0.262 ± 0.061 \\
LaVCa (Ours) & 1 & \ding{55} & 0.267 ± 0.050 & 0.311 ± 0.047 & 0.355 ± 0.054 & 0.241 ± 0.052 \\
\rowcolor{lavcabg} \textbf{LaVCa (Ours)} & 5 & \ding{51} & \textbf{0.314 ± 0.059} & \textbf{0.347 ± 0.053} & \textbf{0.379 ± 0.054} & \textbf{0.289 ± 0.060} \\
\midrule
\multicolumn{7}{c}{\textbf{Top3000 voxels}} \\
\midrule
Model & \# keywords & Sentence Composer & subj01 & subj02 & subj05 & subj07 \\
\midrule
Shuffled & -- & -- & 0.017 ± 0.187 & 0.059 ± 0.210 & 0.087 ± 0.228 & 0.007 ± 0.174 \\
BrainSCUBA & -- & -- & 0.220 ± 0.062 & 0.262 ± 0.063 & 0.291 ± 0.070 & 0.201 ± 0.067 \\
LaVCa (Ours) & 1 & \ding{55} & 0.213 ± 0.058 & 0.255 ± 0.058 & 0.292 ± 0.067 & 0.187 ± 0.061 \\
\rowcolor{lavcabg} \textbf{LaVCa (Ours)} & 5 & \ding{51} & \textbf{0.248 ± 0.066} & \textbf{0.286 ± 0.063} & \textbf{0.315 ± 0.068} & \textbf{0.221 ± 0.069} \\
\midrule
\multicolumn{7}{c}{\textbf{Top5000 voxels}} \\
\midrule
Model & \# keywords & Sentence Composer & subj01 & subj02 & subj05 & subj07 \\
\midrule
Shuffled & -- & -- & 0.017 ± 0.163 & 0.052 ± 0.185 & 0.066 ± 0.204 & 0.009 ± 0.148 \\
BrainSCUBA & -- & -- & 0.188 ± 0.067 & 0.226 ± 0.070 & 0.250 ± 0.078 & 0.169 ± 0.069 \\
LaVCa (Ours) & 1 & \ding{55} & 0.182 ± 0.063 & 0.221 ± 0.066 & 0.252 ± 0.077 & 0.158 ± 0.064 \\
\rowcolor{lavcabg} \textbf{LaVCa (Ours)} & 5 & \ding{51} & \textbf{0.213 ± 0.072} & \textbf{0.249 ± 0.070} & \textbf{0.273 ± 0.079} & \textbf{0.187 ± 0.073} \\
\midrule
\multicolumn{7}{c}{\textbf{Top10000 voxels}} \\
\midrule
Model & \# keywords & Sentence Composer & subj01 & subj02 & subj05 & subj07 \\
\midrule
Shuffled & -- & -- & 0.010 ± 0.128 & 0.034 ± 0.145 & 0.049 ± 0.159 & 0.006 ± 0.114 \\
BrainSCUBA & -- & -- & 0.139 ± 0.073 & 0.170 ± 0.081 & 0.188 ± 0.090 & 0.122 ± 0.073 \\
LaVCa (Ours) & 1 & \ding{55} & 0.134 ± 0.071 & 0.168 ± 0.076 & 0.187 ± 0.091 & 0.114 ± 0.069 \\
\rowcolor{lavcabg} \textbf{LaVCa (Ours)} & 5 & \ding{51} & \textbf{0.160 ± 0.078} & \textbf{0.191 ± 0.082} & \textbf{0.208 ± 0.092} & \textbf{0.138 ± 0.077} \\
\bottomrule
\end{tabular}}
\label{appendix:TopN_image_acc_comparison}
\end{table*}

\clearpage
\begin{figure*}[t] 
  \centering
  \includegraphics[width=\textwidth]{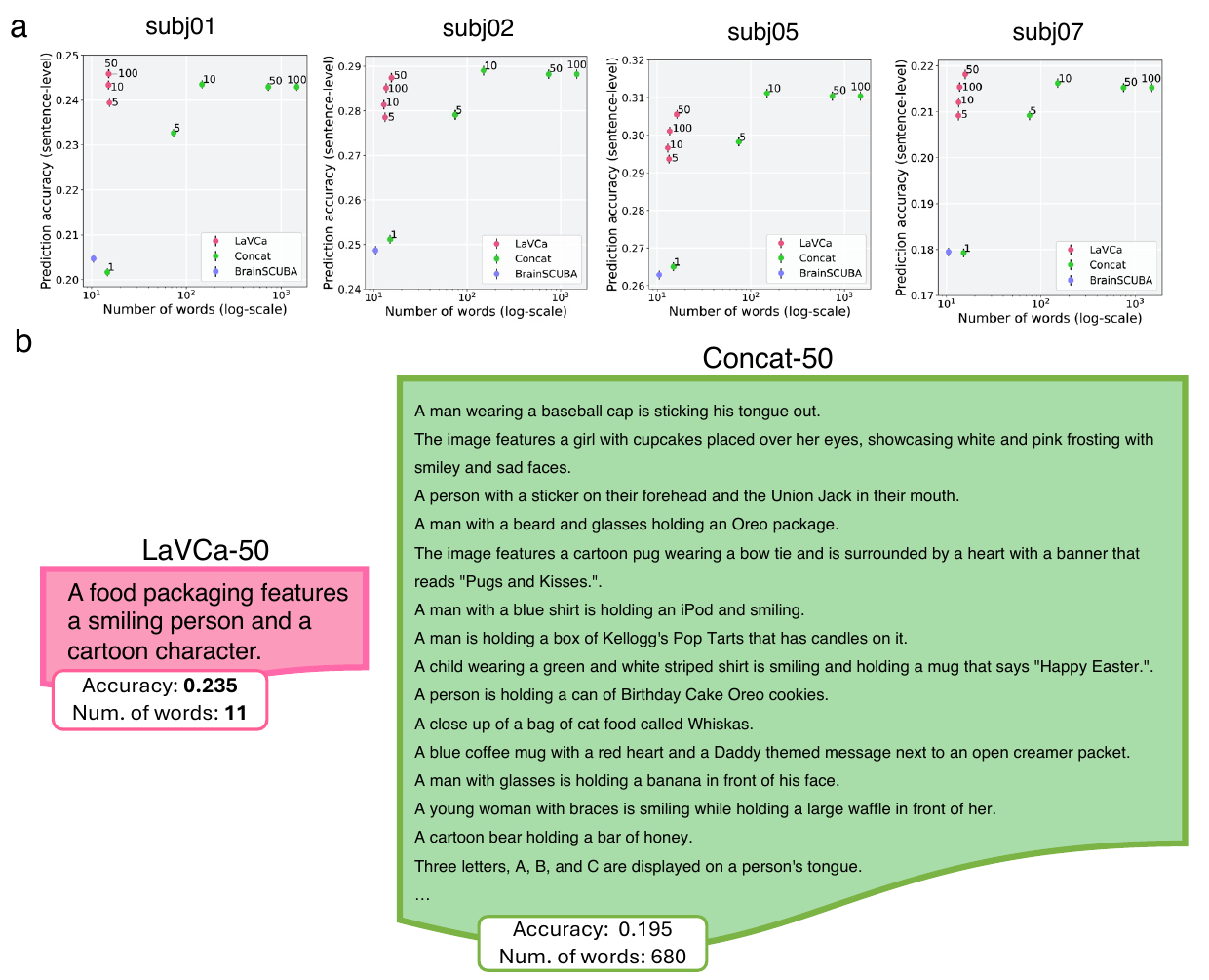}
  \caption{\textbf{a} Relationship between voxel caption prediction performance and word count (all subjects). The color of the plot corresponds to the lineage of each model. The numbers associated with LaVCa indicate the number of optimal images used for summarization, while the numbers associated with Concat represent the number of captions for concatenated optimal images. Error bars indicate the standard error. \textbf{b} Comparison of actual voxel captions between Concat-50 and LaVCa-50. Only a portion of the captions is depicted for Concat-50.
}
  \label{appendix:xaxis_words_all_sub}
\end{figure*}

\clearpage
\begin{table*}[t]
\centering
\caption{Evaluation of the diversity of three models. PCs (90\% Var) means the number of principal components required to explain 90\% variance of the text embeddings. For intra-voxel comparisons, the mean ± standard deviation across subjects is presented. The inter-subject average (Average) is presented as the mean ± standard error.}
\vskip 0.15in
\small 
\scriptsize
\begin{tabular}{@{}llccccccc@{}}
\toprule
 & & & \multicolumn{3}{c}{Inter-voxel} & \multicolumn{3}{c}{Intra-voxel} \\
\cmidrule(lr){4-6}\cmidrule(lr){7-9}
 & & & \multicolumn{1}{c}{Lexical} & \multicolumn{2}{c}{Semantic} & \multicolumn{2}{c}{Lexical} & \multicolumn{1}{c}{Semantic}\\
\cmidrule(lr){4-4}\cmidrule(lr){5-6}\cmidrule(lr){7-8}\cmidrule(lr){9-9}
Subject & Model & Acc. & Vocab. size & Variance & PCs (90\% Var) & Vocab. size & Length & Variance \\ 
\midrule
\multirow{3}{*}{subj01} 
 & BrainSCUBA          &  0.207±0.062 & 3400 & 0.0591 &  99  &  6.21±1.27 &  6.32±1.46 & 0.0163±0.0025 \\
 & Top-1 (Ours)     &  0.202±0.064 & 15384 & 0.0640 & 210 &  9.65±3.59 & 10.0±4.24 & 0.0194±0.0027 \\
 & LaVCa (Ours)        &  0.246±0.066 & 16477 & 0.0639 & 218 & 11.0±2.89 & 11.5±3.19 & 0.0198±0.0025 \\ 
\midrule
\multirow{3}{*}{subj02} 
 & BrainSCUBA          &  0.251±0.071 & 3287 & 0.0591 & 133 &  6.17±1.32 &  6.27±1.51 & 0.0162±0.0026 \\
 & Top-1 (Ours)     &  0.251±0.070 & 14135 & 0.0632 & 206 &  9.99±3.65 & 10.5±4.28 & 0.0195±0.0027 \\
 & LaVCa (Ours)        &  0.287±0.075 & 17242 & 0.0639 & 218 & 11.3±3.64 & 11.8±3.93 & 0.0198±0.0027 \\ 
\midrule
\multirow{3}{*}{subj05} 
 & BrainSCUBA          &  0.263±0.084 & 3043 & 0.0583 & 127 &  6.18±1.37 &  6.26±1.52 & 0.0159±0.0027 \\
 & Top-1 (Ours)     &  0.265±0.081 & 13485 & 0.0631 & 206 &  9.99±3.68 & 10.4±4.34 & 0.0195±0.0028 \\
 & LaVCa (Ours)        &  0.306±0.084 & 17459 & 0.0644 & 218 & 11.8±3.88 & 12.2±4.14 & 0.0199±0.0027 \\ 
\midrule
\multirow{3}{*}{subj07} 
 & BrainSCUBA          &  0.182±0.065 & 3042 & 0.0587 & 131 &  6.23±1.30 &  6.36±1.47 & 0.0163±0.0026 \\
 & Top-1 (Ours)     &  0.179±0.066 & 12831 & 0.0632 & 203 & 10.1±3.51 & 10.6±4.14 & 0.0197±0.0026 \\
 & LaVCa (Ours)        &  0.218±0.073 & 16508 & 0.0646 & 222 & 11.6±3.76 & 12.0±4.02 & 0.0202±0.0026 \\
\midrule
\multirow{3}{*}{Average} 
 & BrainSCUBA          
   & 0.226±0.019 & 3193±90 & 0.0588±0.0002 & 123±7.93 
   & 6.20±0.01 & 6.30±0.02 & 0.0162±0.0001 \\
 & Top-1 (Ours) 
   & 0.224±0.020 & 13959±545 & 0.0634±0.0002 & 206±1.44 
   & 9.93±0.10 & 10.4±0.132 & 0.0195±0.0001 \\
 & LaVCa (Ours) 
   & 0.264±0.020 & 16922±252 & 0.0642±0.0002 & 219±1.00 
   & 11.4±0.175 & 11.9±0.149 & 0.0199±0.0001 \\
\bottomrule
\end{tabular}
\label{table:diversity_analysis}
\end{table*}

\clearpage

\begin{table*}[t]
\centering
\caption{The average prediction accuracy for each subject and the inter-subject average prediction accuracy when captions were shuffled within the ROI (Shuffled) and when they were used as-is (Original). For each subject, the average prediction accuracy ± standard deviation is depicted, while for the inter-subject average, the average prediction accuracy ± standard error is presented.}
\vskip 0.15in
{
\begin{tabular}{@{}cc*{4}{r}@{}}
\toprule
& & \multicolumn{2}{c}{\textbf{Body areas}} & \multicolumn{2}{c}{\textbf{Face areas}} \\
\cmidrule(lr){3-4}\cmidrule(lr){5-6}
& & EBA & FBA-2 & OFA & FFA-1 \\
\midrule
\multirow[c]{2}{*}{subj01} & Shuffled & 0.035±0.147 & 0.014±0.128 & 0.031±0.067 & 0.017±0.113 \\
& \textbf{Original} & \textbf{0.169±0.105} & \textbf{0.124±0.102} & \textbf{0.083±0.069} & \textbf{0.117±0.083} \\
\cline{1-6}
\multirow[c]{2}{*}{subj02} & Shuffled & 0.010±0.144 & 0.026±0.109 & 0.036±0.071 & 0.024±0.097 \\
& \textbf{Original} & \textbf{0.158±0.101} & \textbf{0.128±0.103} & \textbf{0.079±0.066} & \textbf{0.105±0.078} \\
\cline{1-6}
\multirow[c]{2}{*}{subj05} & Shuffled & -0.001±0.148 & 0.007±0.148 & 0.017±0.118 & 0.009±0.116 \\
& \textbf{Original} & \textbf{0.152±0.111} & \textbf{0.149±0.114} & \textbf{0.120±0.100} & \textbf{0.112±0.089} \\
\cline{1-6}
\multirow[c]{2}{*}{subj07} & Shuffled & 0.028±0.135 & 0.025±0.096 & 0.027±0.096 & 0.013±0.100 \\
& \textbf{Original} & \textbf{0.149±0.104} & \textbf{0.099±0.099} & \textbf{0.097±0.099} & \textbf{0.108±0.090} \\
\cline{1-6}
\multirow[c]{2}{*}{Average} & Shuffled & 0.018±0.008 & 0.018±0.005 & 0.028±0.004 & 0.016±0.003 \\
& \textbf{Original} & \textbf{0.157±0.005} & \textbf{0.125±0.010} & \textbf{0.095±0.009} & \textbf{0.111±0.003} \\
\midrule
& & \multicolumn{2}{c}{\textbf{Place areas}} & \multicolumn{2}{c}{\textbf{Word areas}} \\
\cmidrule(lr){3-4}\cmidrule(lr){5-6}
& & OPA & PPA & OWFA & VWFA-1 \\
\midrule
\multirow[c]{2}{*}{subj01} & Shuffled & 0.080±0.108 & 0.105±0.107 & 0.015±0.057 & 0.054±0.114 \\
& \textbf{Original} & \textbf{0.163±0.093} & \textbf{0.172±0.099} & \textbf{0.055±0.048} & \textbf{0.147±0.088} \\
\cline{1-6}
\multirow[c]{2}{*}{subj02} & Shuffled & 0.118±0.139 & 0.178±0.147 & 0.037±0.071 & 0.031±0.135 \\
& \textbf{Original} & \textbf{0.204±0.114} & \textbf{0.243±0.139} & \textbf{0.085±0.066} & \textbf{0.150±0.099} \\
\cline{1-6}
\multirow[c]{2}{*}{subj05} & Shuffled & 0.184±0.140 & 0.217±0.153 & 0.028±0.109 & 0.039±0.148 \\
& \textbf{Original} & \textbf{0.260±0.124} & \textbf{0.275±0.149} & \textbf{0.118±0.105} & \textbf{0.177±0.108} \\
\cline{1-6}
\multirow[c]{2}{*}{subj07} & Shuffled & 0.083±0.119 & 0.105±0.108 & 0.020±0.070 & 0.012±0.159 \\
& \textbf{Original} & \textbf{0.175±0.096} & \textbf{0.163±0.106} & \textbf{0.079±0.069} & \textbf{0.157±0.112} \\
\cline{1-6}
\multirow[c]{2}{*}{Average} & Shuffled & 0.116±0.024 & 0.151±0.028 & 0.025±0.005 & 0.034±0.009 \\
& \textbf{Original} & \textbf{0.200±0.022} & \textbf{0.213±0.027} & \textbf{0.084±0.013} & \textbf{0.158±0.007} \\
\bottomrule
\end{tabular}
}
\end{table*}

\clearpage
\begin{figure*}[t]
    \centering
    \includegraphics[width=\textwidth]{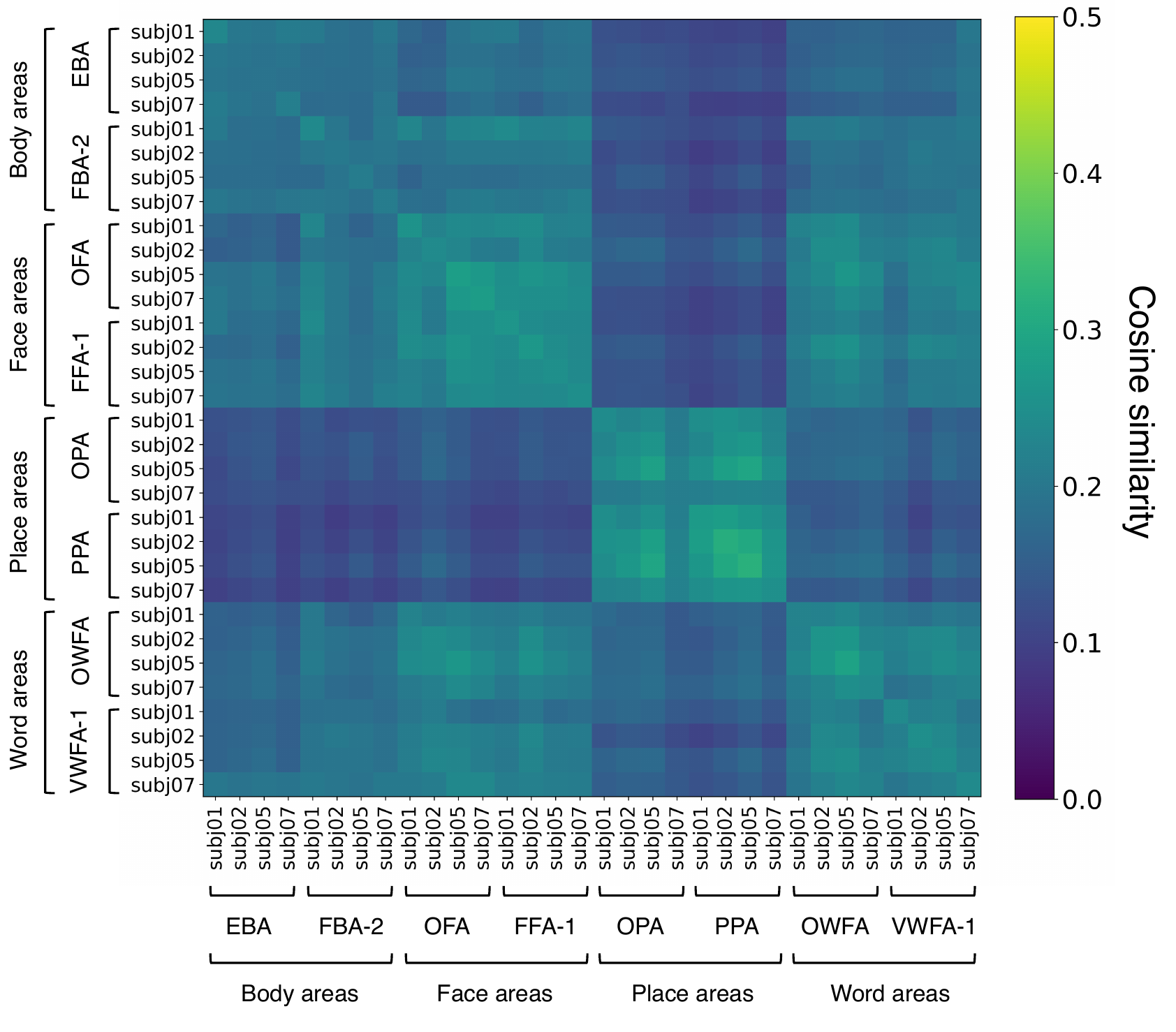}
    \caption{Cross‐subject similarity of voxel captions in ROIs. Each cell shows the mean cosine similarity between sentence embeddings of all voxel captions in the two sets.}
    \label{results:inter-roi_inter-subj}
\end{figure*}

\begin{table*}[t]
\centering
\caption{%
Voxel counts for each ROI, categorized by the number of clusters to which each voxel belongs (\emph{\# assigned clusters}). Average rows show the mean $\pm$ standard error across subjects.}
\vspace{1em}
\label{table:multi_concept}
\scriptsize               
\setlength{\tabcolsep}{3pt} 

\resizebox{0.9\textwidth}{!}{
\begin{tabular}{llrrrrrrrr}
\toprule
 &  & \multicolumn{2}{c}{Body areas}
    & \multicolumn{2}{c}{Face areas}
    & \multicolumn{2}{c}{Place areas}
    & \multicolumn{2}{c}{Word areas}\\
\cmidrule(lr){3-4}\cmidrule(lr){5-6}\cmidrule(lr){7-8}\cmidrule(lr){9-10}
 & \# assigned clusters & EBA & FBA-2 & OFA & FFA-1 & OPA & PPA & OWFA & VWFA-1\\
\midrule
\multirow{6}{*}{\textbf{subj01}}
 & 1 & 40 & 11 & 10 & 14 & 14 & 14 &  4 &  6\\
 & 2 & 517 & 45 & 39 & 60 & 141 & 90 & 60 & 74\\
 & 3 & 1083 & 126 & 130 & 151 & 465 & 288 & 136 & 257\\
 & 4 & 904 & 147 & 112 & 165 & 579 & 369 & 165 & 263\\
 & 5 & 377 & 82 & 53 & 84 & 325 & 226 & 83 & 149\\
 & 6 & 50 & 19 & 11 & 10 & 87 & 46 & 16 & 23\\
\midrule
\multirow{6}{*}{\textbf{subj02}}
 & 1 & 46 & 13 &  4 &  5 & 20 & 16 &  3 &  2\\
 & 2 & 517 & 137 & 49 & 44 & 182 & 140 & 39 & 32\\
 & 3 & 1037 & 381 & 124 & 124 & 457 & 380 & 137 & 108\\
 & 4 & 1081 & 424 & 147 & 112 & 454 & 337 & 207 & 124\\
 & 5 & 624 & 219 & 95 & 52 & 216 & 116 & 106 & 64\\
 & 6 & 134 & 43 & 22 &  3 & 52 &  5 & 27 & 15\\
\midrule
\multirow{6}{*}{\textbf{subj05}}
 & 1 & 39 & 7 & 9 & 29 & 19 & 13 & 5 &  8\\
 & 2 & 607 & 72 & 102 & 108 & 178 & 171 & 62 & 55\\
 & 3 & 1449 & 182 & 257 & 160 & 416 & 444 & 134 & 139\\
 & 4 & 1446 & 181 & 248 & 111 & 450 & 379 & 147 & 157\\
 & 5 & 829 & 65 & 139 & 41 & 218 & 177 & 73 & 92\\
 & 6 & 214 & 1 & 26 & 3 & 50 & 37 & 17 & 35\\
\midrule
\multirow{6}{*}{\textbf{subj07}}
 & 1 & 34 & 11 &  4 &  5 & 34 &  8 &  1 &  6\\
 & 2 & 303 & 69 & 26 & 45 & 193 & 118 & 40 & 34\\
 & 3 & 1123 & 158 & 88 & 88 & 393 & 298 & 134 & 112\\
 & 4 & 1041 & 180 & 99 & 113 & 308 & 323 & 267 & 101\\
 & 5 & 492 & 116 & 84 & 72 & 141 & 143 & 155 & 67\\
 & 6 & 69 & 18 & 15 & 23 & 14 & 22 & 31 & 7\\
\midrule
\multirow{6}{*}{\textbf{Average}}
 & 1 & 40$\pm$2 & 10$\pm$1 &  7$\pm$2 & 13$\pm$6 & 22$\pm$4 & 13$\pm$2 &  3$\pm$1 &  6$\pm$1\\
 & 2 & 486$\pm$65 & 81$\pm$20 & 54$\pm$17 & 64$\pm$15 & 174$\pm$11 & 130$\pm$17 & 50$\pm$6 & 49$\pm$10\\
 & 3 & 1173$\pm$94 & 212$\pm$58 & 150$\pm$37 & 131$\pm$16 & 433$\pm$17 & 352$\pm$37 & 135$\pm$1 & 154$\pm$35\\
 & 4 & 1118$\pm$116 & 233$\pm$64 & 152$\pm$34 & 125$\pm$13 & 448$\pm$55 & 352$\pm$13 & 197$\pm$27 & 161$\pm$36\\
 & 5 & 580$\pm$97 & 120$\pm$35 & 93$\pm$18 & 62$\pm$10 & 225$\pm$38 & 166$\pm$24 & 104$\pm$18 & 93$\pm$20\\
 & 6 & 117$\pm$37 & 20$\pm$9 & 18$\pm$3 & 10$\pm$5 & 51$\pm$15 & 28$\pm$9 & 23$\pm$4 & 20$\pm$6\\
\bottomrule
\end{tabular}} 
\end{table*}


\clearpage
\begin{figure*}[t]
    \centering
    \includegraphics[width=\textwidth]{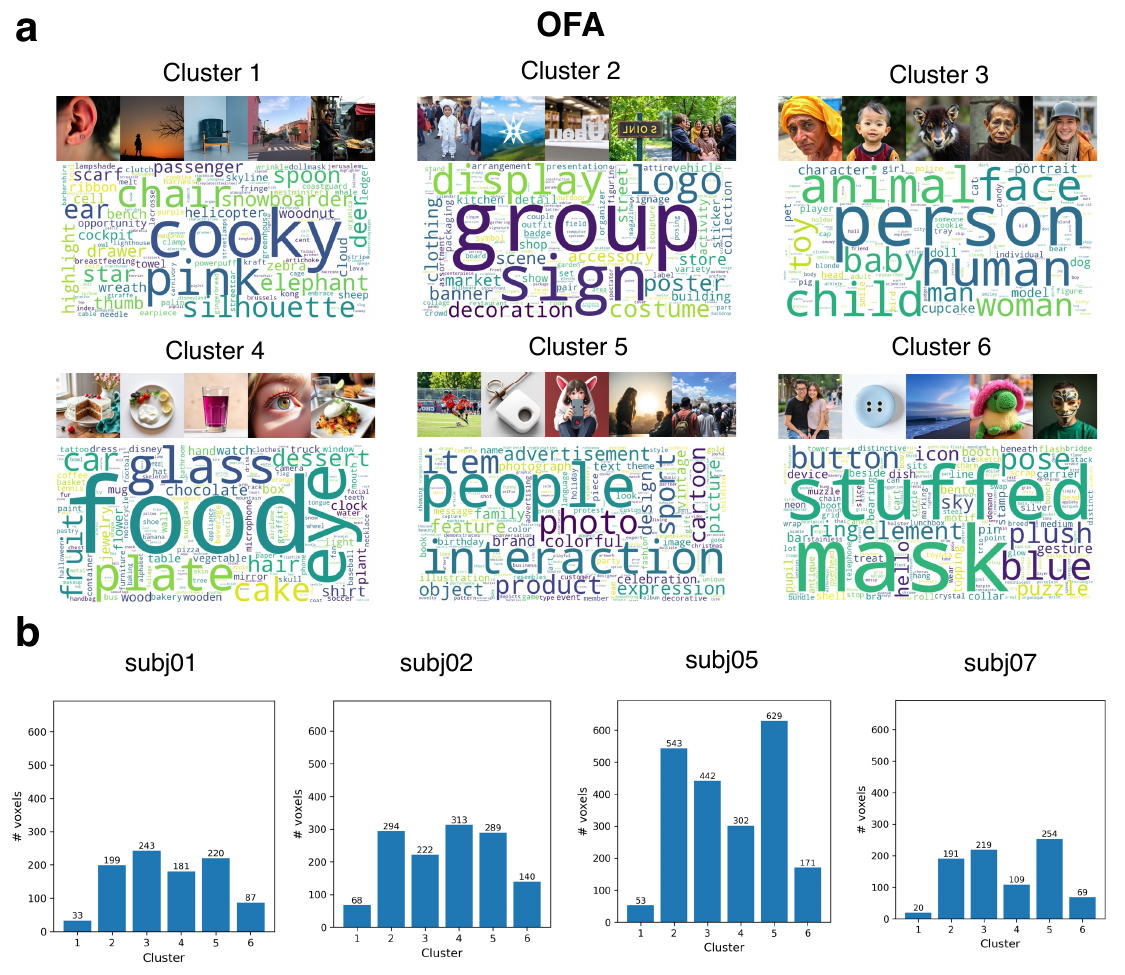}
    \caption{Subject-shared noun cluster analysis in the OFA. \textbf{a} Word clouds and generated images for the top-5 most frequent nouns in each subject-shared cluster. \textbf{b} Bar graphs showing the number of voxels assigned to each subject-shared cluster for individual subjects.}
    \label{appendix:inter-subject-cluster-anlysis}
\end{figure*} 

\clearpage
\begin{figure*}[t]
    \centering
    \includegraphics[width=\textwidth]{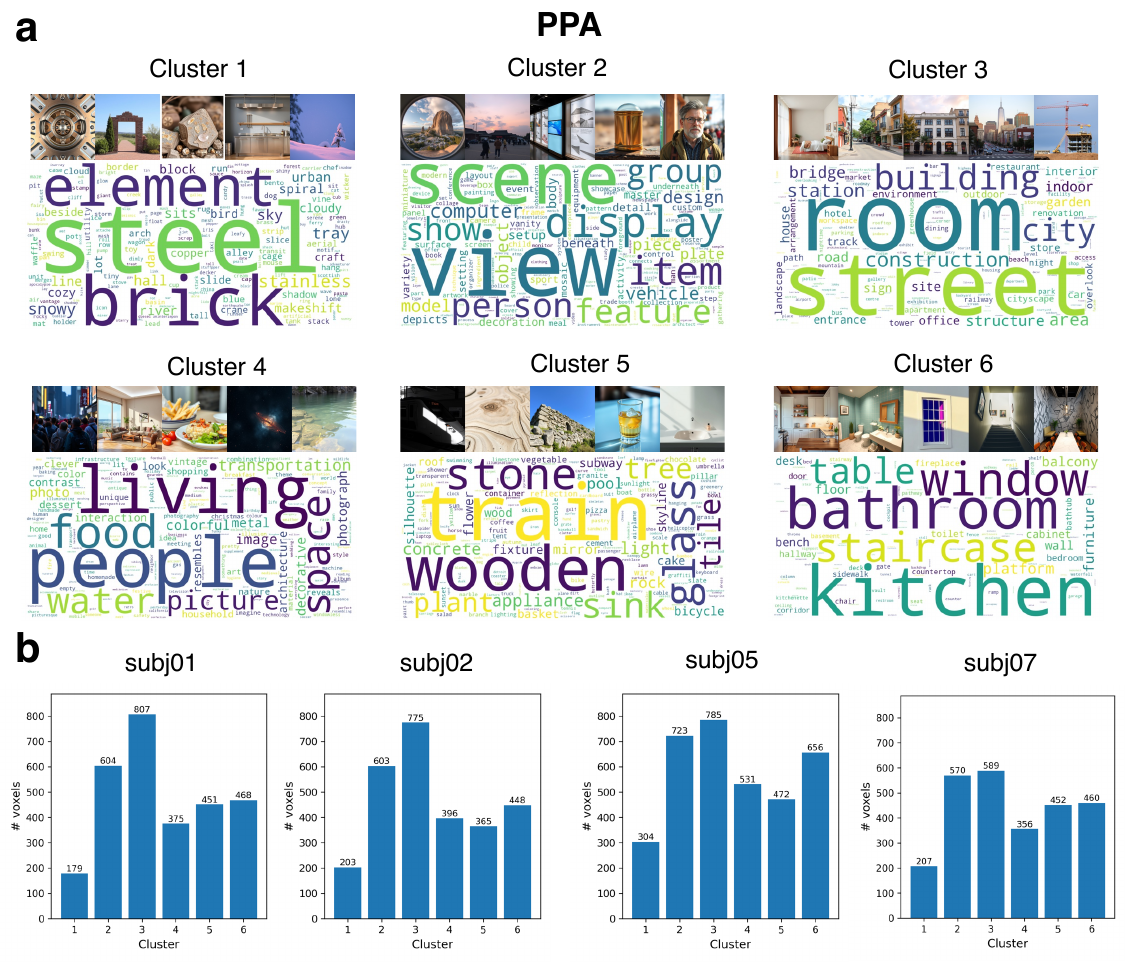}
    \caption{Subject-shared noun cluster analysis in the PPA. \textbf{a} Word clouds and generated images for the top-5 most frequent nouns in each subject-shared cluster. \textbf{b} Bar graphs showing the number of voxels assigned to each subject-shared cluster for individual subjects.}
\end{figure*}


\clearpage
\begin{figure*}[ht] 
  \centering
  \includegraphics[width=\textwidth]{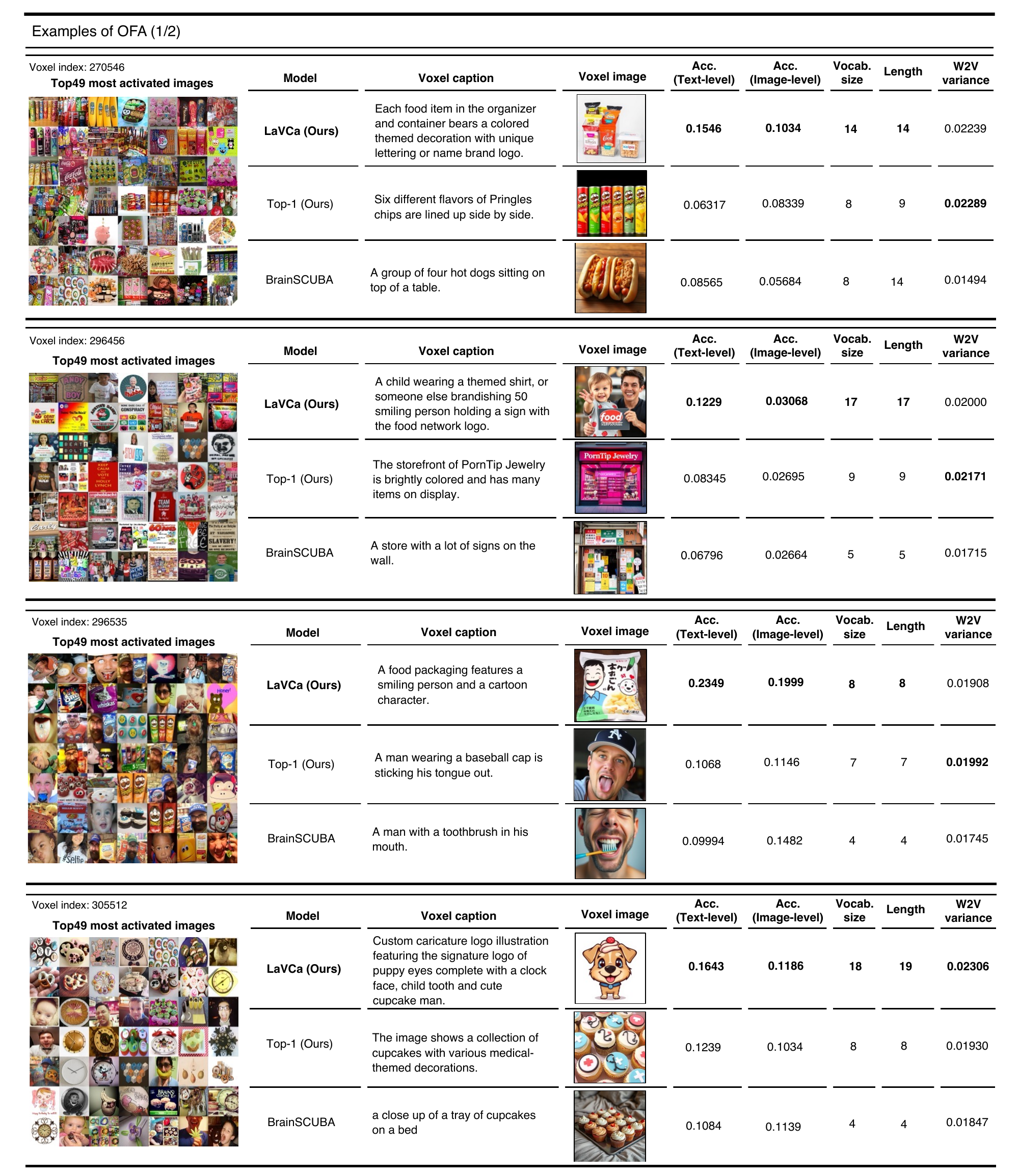}
  \caption{Comparison of voxel captions and voxel images in the OFA voxels of subj02 (1/2).
}
  \label{appendix:OFA_example_1}
\end{figure*}

\clearpage
\begin{figure*}[t] 
  \centering
  \includegraphics[width=\textwidth]{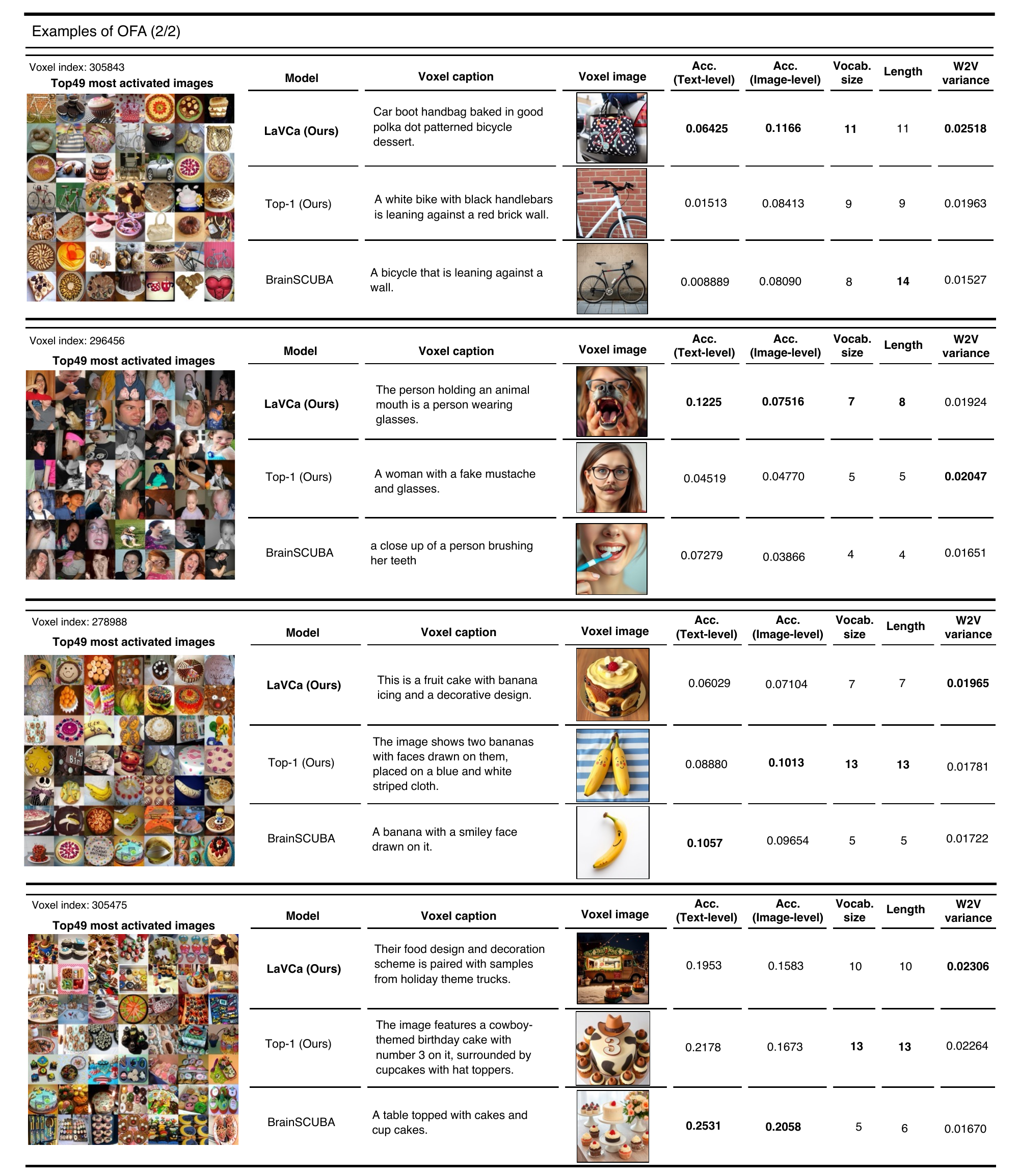}
  \caption{Comparison of voxel captions and voxel images in the OFA voxels of subj02 (2/2).
}
  \label{appendix:OFA_example_2}
\end{figure*}

\clearpage
\begin{figure*}[t] 
  \centering
  \includegraphics[width=\textwidth]{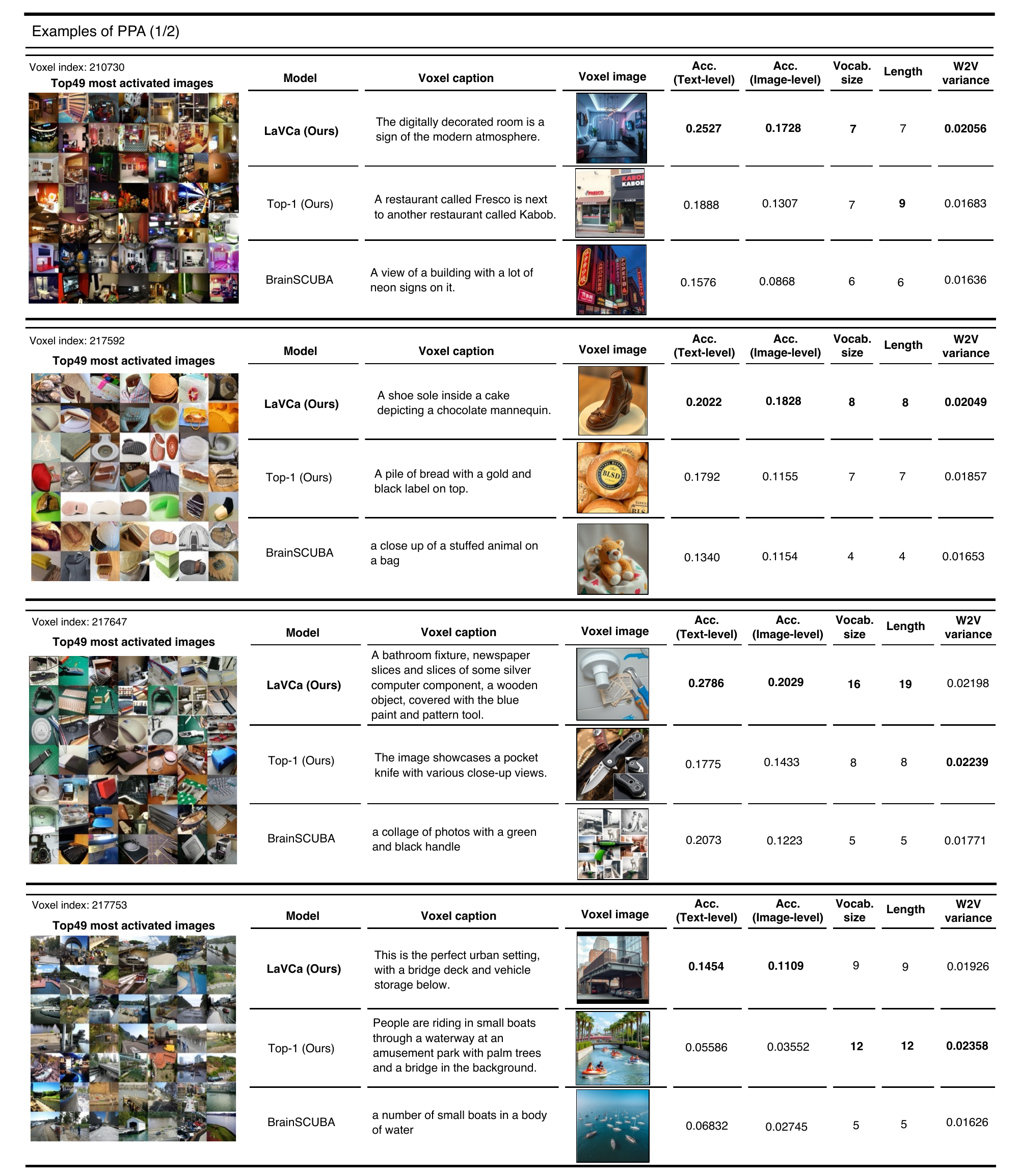}
  \caption{Comparison of voxel captions and voxel images in the PPA voxels of subj07 (1/2).
}
  \label{appendix:PPA_example_1}
\end{figure*}

\clearpage
\begin{figure*}[t] 
  \centering
  \includegraphics[width=\textwidth]{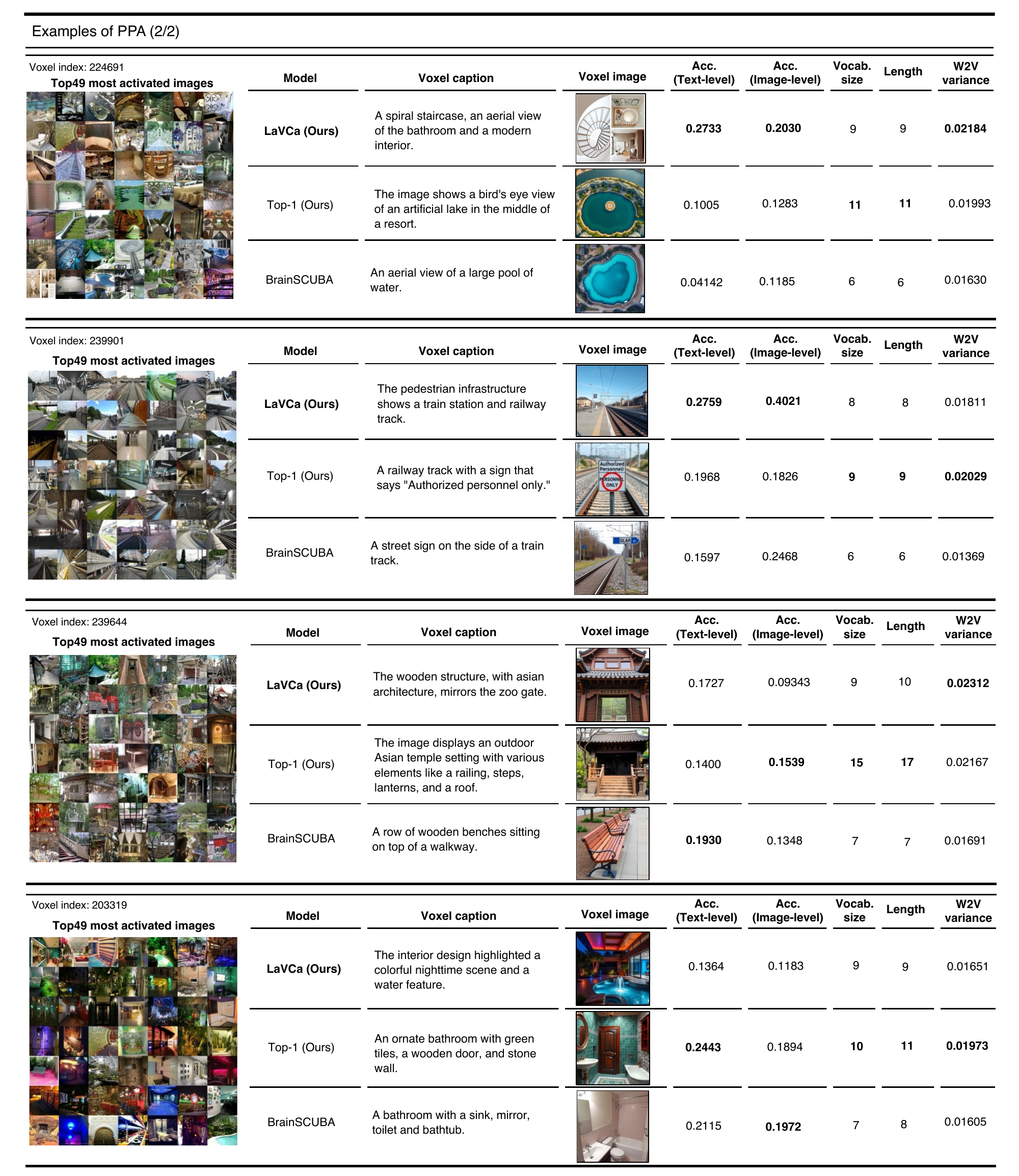}
  \caption{Comparison of voxel captions and voxel images in the PPA voxels of subj07 (2/2).
}
  \label{appendix:PPA_example_2}
\end{figure*}

\clearpage
\begin{figure*}[t] 
  \centering
  \includegraphics[width=\textwidth]{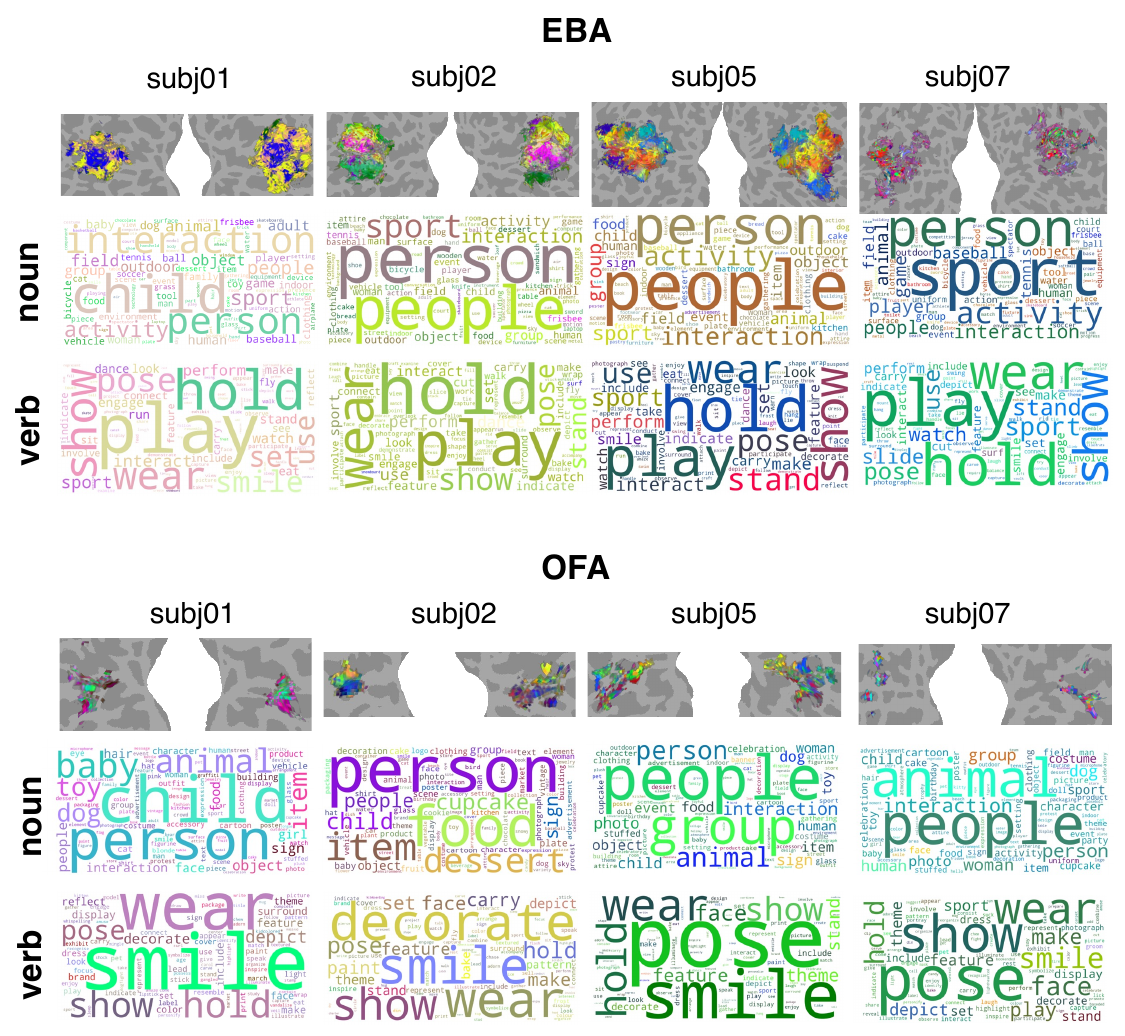}
  \caption{The UMAP projection of caption text across EBA and OFA for all subjects, visualized on a flatmap (top). 
    A word cloud of the 100 most frequent \textbf{nouns} in these captions (middle), colored according to their location in the UMAP space. 
    A word cloud of the 100 most frequent \textbf{verbs} (bottom). 
}
  \label{appendix:EBA_OFA_WordCloud}
\end{figure*}

\clearpage
\begin{figure*}[t] 
  \centering
  \includegraphics[width=\textwidth]{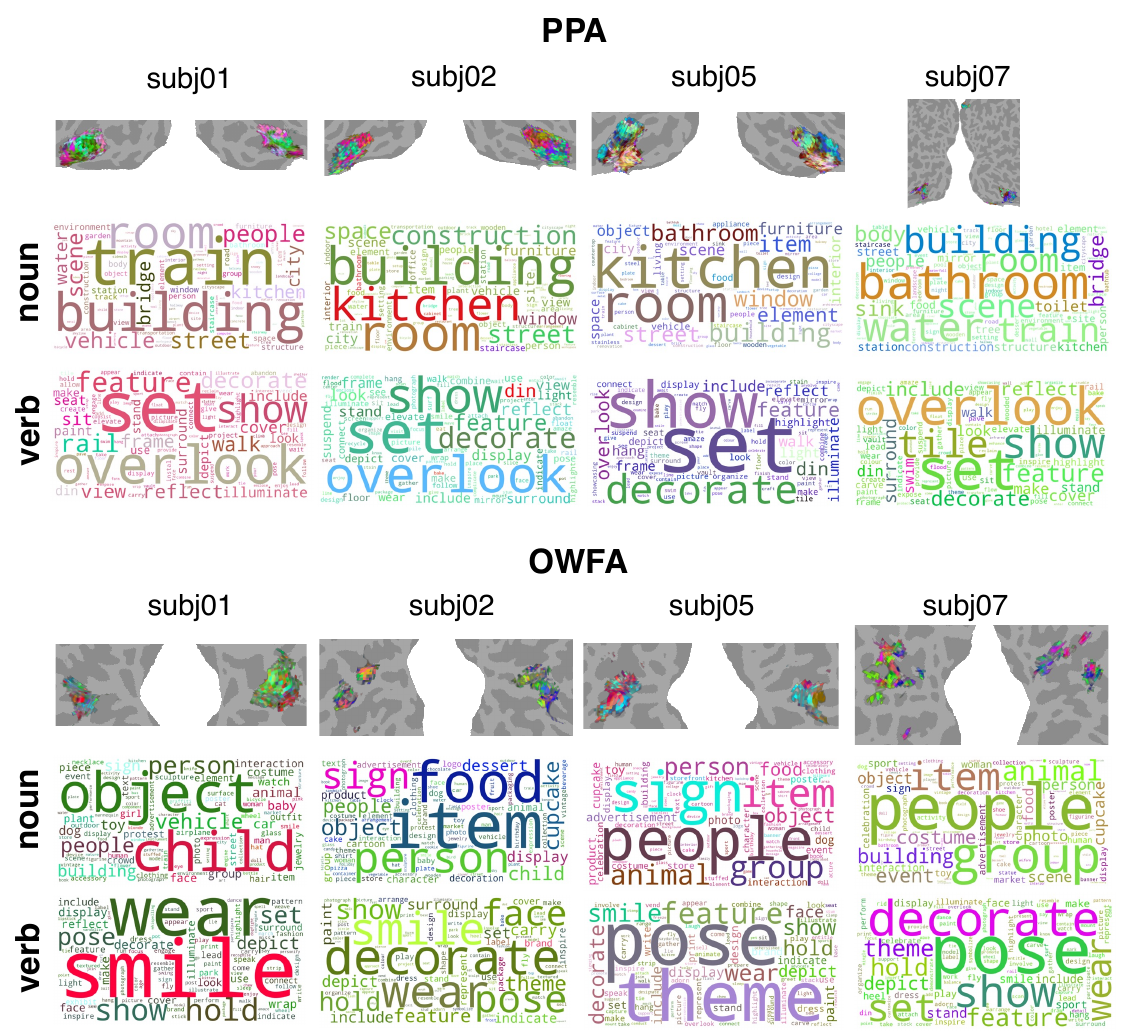}
  \caption{The UMAP projection of caption text across PPA and OWFA for all subjects, visualized on a flatmap (top). 
    A word cloud of the 100 most frequent \textbf{nouns} in these captions (middle), colored according to their location in the UMAP space. 
    A word cloud of the 100 most frequent \textbf{verbs} (bottom). 
}
  \label{appendix:PPA_OWFA_WordCloud}
\end{figure*}

\clearpage
\begin{figure*}[t] 
  \centering
  \includegraphics[width=\textwidth]{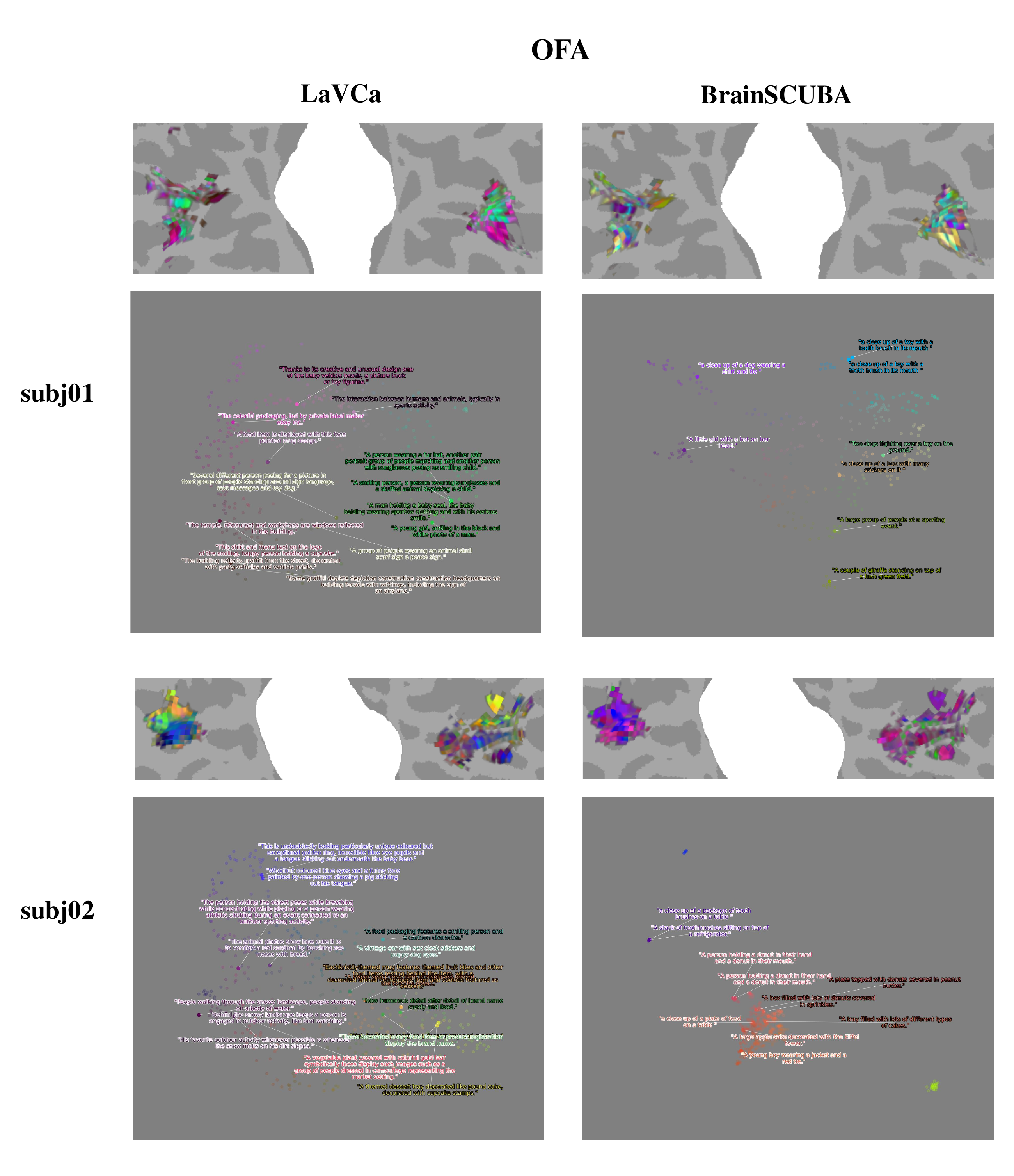}
  \caption{Visualization of OFA captions for subj01 and subj02. For each subject, the captions' UMAP representations were mapped onto a flatmap (top). The top 2 captions of each cluster in the UMAP space were visualized (bottom). The horizontal axis represents UMAP2, and the vertical axis represents UMAP2.
}
  \label{appendix:OFA_caps_umap_subj01-02}
\end{figure*}

\clearpage
\begin{figure*}[t] 
  \centering
  \includegraphics[width=\textwidth]{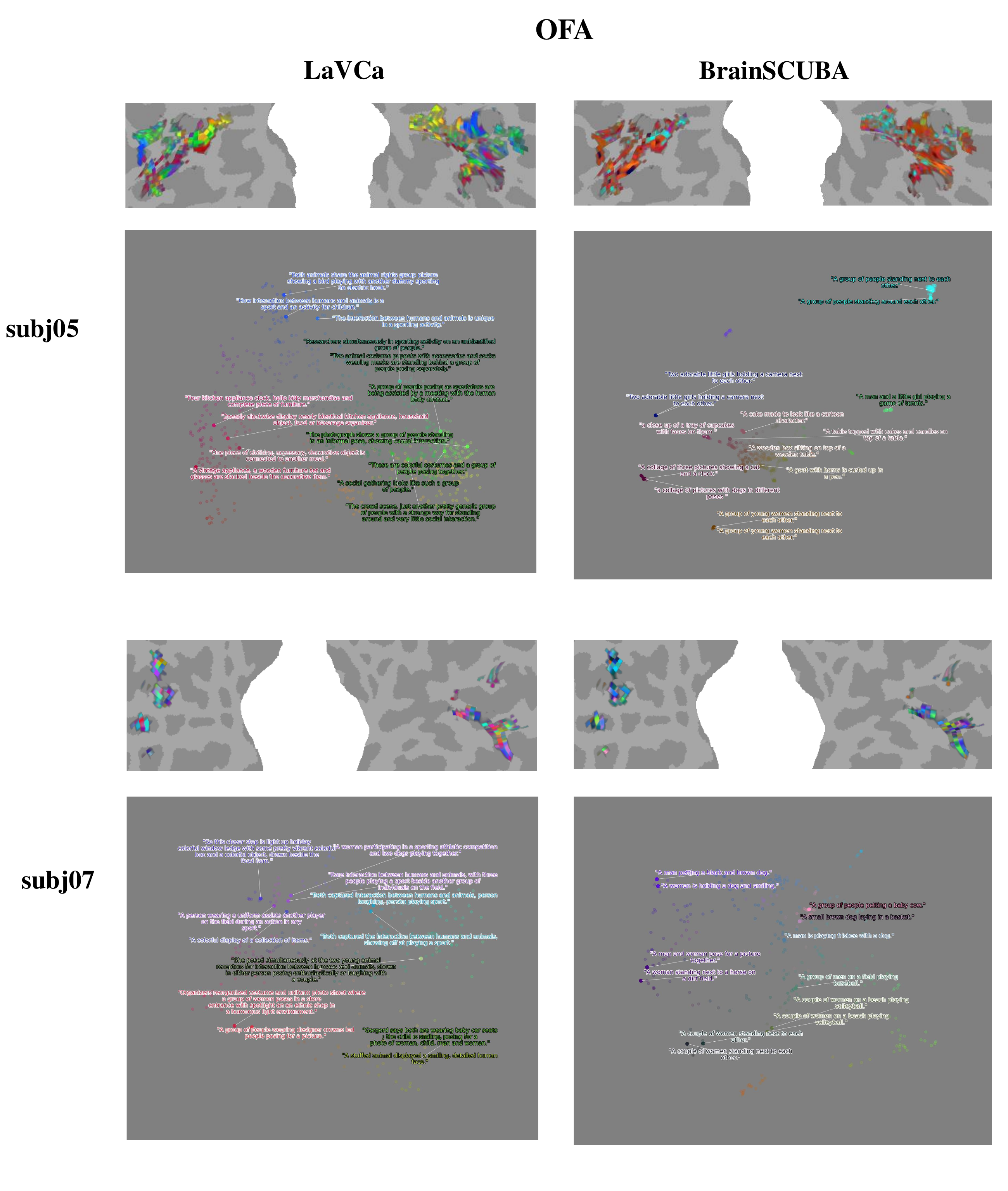}
  \caption{Visualization of OFA captions for subj05 and subj07. The captions’ UMAP representations were mapped onto a flatmap (top) for each subject. The top 2 captions of each cluster in the UMAP space were visualized (bottom). The horizontal axis represents UMAP2, and the vertical axis represents UMAP2.
}
  \label{appendix:OFA_caps_umap_subj05-07}
\end{figure*}

\clearpage
\begin{figure*}[t] 
  \centering
  \includegraphics[width=\textwidth]{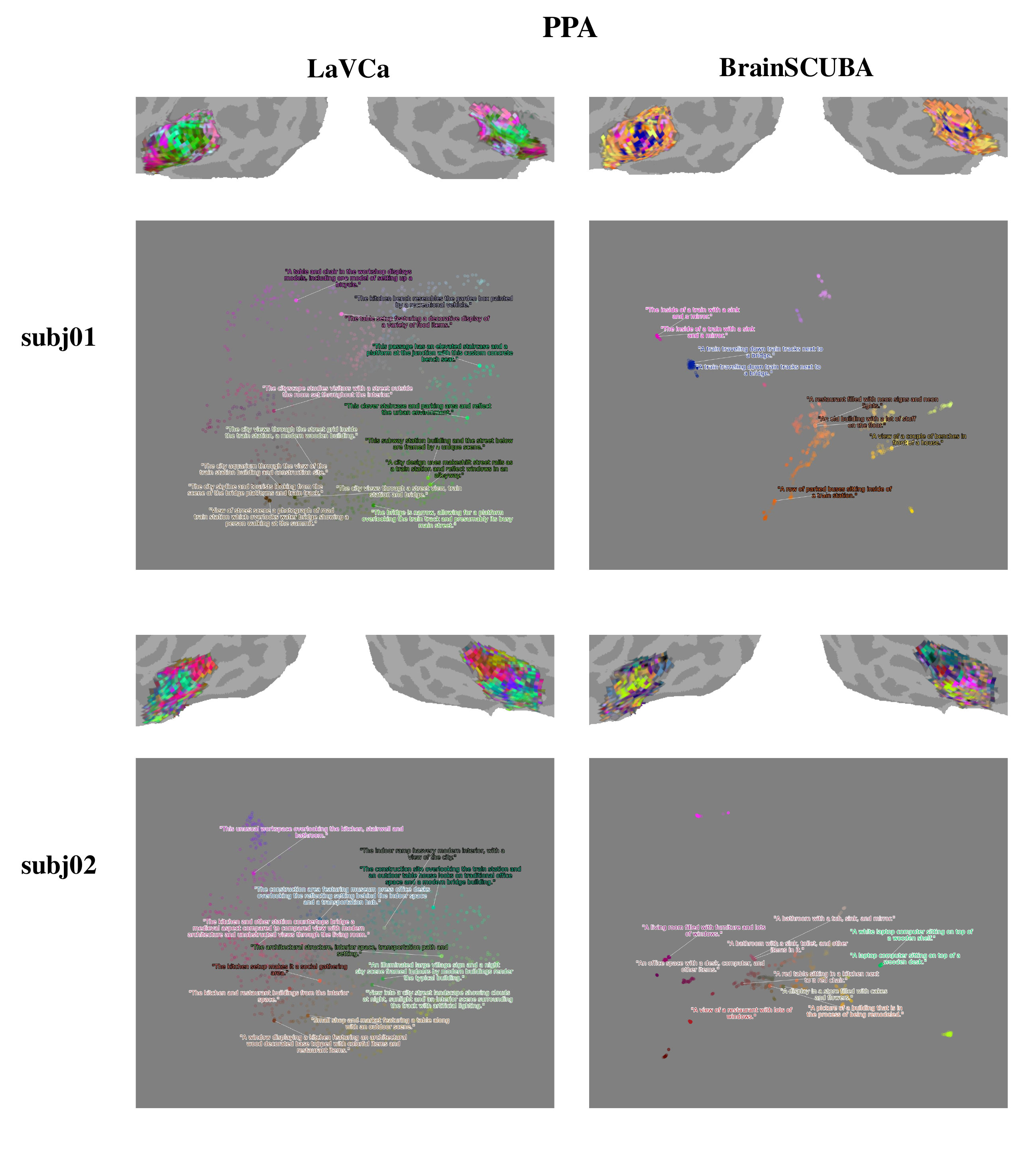}
  \caption{Visualization of PPA captions for subj01 and subj02. The captions’ UMAP representations were mapped onto a flatmap (top) for each subject. The top 2 captions of each cluster in the UMAP space were visualized (bottom). The horizontal axis represents UMAP2, and the vertical axis represents UMAP2.
}
  \label{appendix:PPA_caps_umap_subj01-02}
\end{figure*}

\clearpage
\begin{figure*}[t] 
  \centering
  \includegraphics[width=0.95\textwidth]{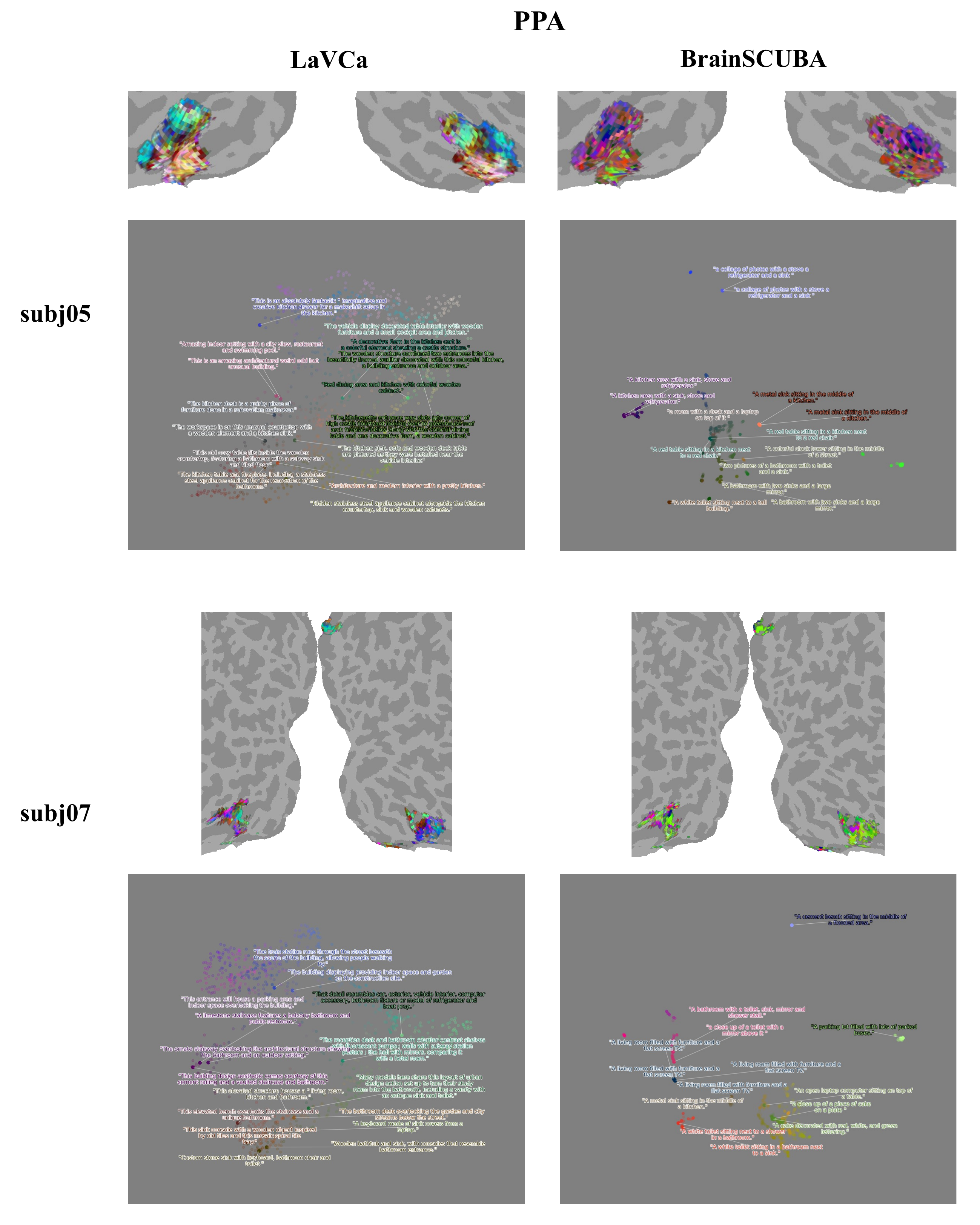}
  \caption{Visualization of PPA captions for subj05 and subj07. The captions’ UMAP representations were mapped onto a flatmap (top) for each subject. The top 2 captions of each cluster in the UMAP space were visualized (bottom). The horizontal axis represents UMAP2, and the vertical axis represents UMAP2.
}
  \label{appendix:PPA_caps_umap_subj05-07}
\end{figure*}

\end{document}